\documentclass[%
 aip,
 amsmath,amssymb,
 reprint,%
]{revtex4-1}

\usepackage{graphicx}% Include figure files
\usepackage{dcolumn}% Align table columns on decimal point
\usepackage{bm}% bold math

\usepackage[utf8]{inputenc}
\usepackage[T1]{fontenc}
\usepackage{etoolbox}
\usepackage{txfonts}

\usepackage{color}

\makeatletter
\def\@email#1#2{%
 \endgroup
 \patchcmd{\titleblock@produce}
  {\frontmatter@RRAPformat}
  {\frontmatter@RRAPformat{\produce@RRAP{*#1\href{mailto:#2}{#2}}}\frontmatter@RRAPformat}
  {}{}
}%
\makeatother
\begin{document}

\preprint{AIP/123-QED}

\title[BOnd TArgeting Network for predicting glassy dynamics]{BOTAN: BOnd TArgeting Network for prediction of slow glassy dynamics by machine learning relative motion}
% Force line breaks with \\

\author{Hayato Shiba}
\email{shiba@cc.u-tokyo.ac.jp}
\affiliation{Information Technology Center, University of Tokyo, Chiba 277-0882, Japan}
\author{Masatoshi Hanai}%
\affiliation{Information Technology Center, University of Tokyo, Chiba 277-0882, Japan}
\author{Toyotaro Suzumura}
\affiliation{Graduate School of Information Science and Technology, University of Tokyo, Tokyo 133-8658, Japan}
\affiliation{Information Technology Center, University of Tokyo, Chiba 277-0882, Japan}
\author{Takashi Shimokawabe}
%\email{E-mail: shimokawabe@cc.u-tokyo.ac.jp}
\affiliation{Information Technology Center, University of Tokyo, Chiba 277-0882, Japan}

\date{\today}% It is always \today, today,

\begin{abstract}
Recent developments in machine learning have enabled accurate predictions of the dynamics of slow structural relaxation in glass-forming systems. However, existing machine-learning models for these tasks are mostly designed such that they learn a single dynamic quantity and relate it to the structural features of glassy liquids. In this study, we propose a graph neural network model, ``BOnd TArgeting Network (BOTAN)'', that learns relative motion between neighboring pairs of particles, in addition to the self-motion of particles. By relating the structural features to these two different dynamical variables, the model autonomously acquires the ability to discern how different dynamical processes, strain fluctuations and particle rearrangements, affect the self-motion of particles undergoing slow relaxation, and thus can predict with high precision how slow structural relaxation develops in space and time.
\end{abstract}

\maketitle

\section{\label{sec:intro} Introduction}
When a liquid is rapidly cooled, it freezes into a glass which retains a random amorphous structure.  This phenomenon, referred to as the glass transition, is ubiquitous, but its origin still remains to be clarified. The elucidation of this origin has been an important research topic for decades~\cite{Angell1995,Debenedetti2001}. Along with the drastic slowdown of atomic motion accompanying the glass transition, localized domains of particles that rearrange more preferentially than others grow up~\cite{Ediger2000,Kob1997,Yamamoto1998b,Berthier2011}.
Identifying the structural origin of this phenomenon has long been a central problem in the field. Structural analyses of  local geometric orders~\cite{Tanaka2019,Royall2015,Hua2018}
and of real-space normal modes of vibrations derived from static structures~\cite{Harrowell2008,Tanguy2010,Yodh2011,Bonn2011,Schoenholtz2014} have accumulated evidence for structure--dynamics correspondence. However, the random, featureless structures of glasses still prevent us from fully understanding the origin and predicting the dynamics based on the static structure. 

Recently, machine-learning methodologies have enabled the accurate determination of structure--dynamics correspondence in glassy systems~\cite{Nussinov2011,Cubuk2015,Schoenholtz2016,Schoenholtz2017,Filion2020,Bapst2020,Filion2021,Filion2022,Coslovich2022}. Among them, graph neural networks (GNNs)~\cite{Scarselli2009}, a type of deep-learning algorithm that operates on a graph, have been found to be suitable~\cite{Bapst2020}. 
GNNs had already proven to be useful for  making predictions of material properties by directly representing  structures of molecules or atomic clusters as nodes and edges of the graph~\cite{CGCNN_2018,OpenCatalyst2021,Fung2021,PFN2022}, Bapst {\it et al.}~\cite{Bapst2020} applied this ability of GNNs to the task of predicting particle mobility in a glass-forming liquid from its static structure, and showed that GNNs outperform other machine-learning methodologies. This high accuracy in the predictions of glassy dynamics has enabled the development of inverse modeling as a further application~\cite{Wang2021}. It has also led to the development of a cost-effective linear-regression model that can overcome the high computational cost of GNNs but still achieves equivalent precision in its prediction~\cite{Filion2021,Filion2022}.

In the study by Bapst {\it et al.}~\cite{Bapst2020}, the prediction accuracy of GNNs was found to exceed not only that of other machine-learning methods but also that of normal mode analysis, at all temperatures accessible by simulations and across all temporal scales up to the structural relaxation time. Normal mode analysis is a method to predict the spatial distribution of short-time vibrations.  Because short-time vibrations (soft modes) are  intimately linked to the slow $\alpha$  relaxations~\cite{Buchenau1992,Widmer-Cooper2006,Dyre2006,Leporini2016,Dyre2017,Hua2018}, normal mode analysis yields a strong baseline of physics-based prediction of glassy dynamics from a static structure~\cite{Harrowell2008,Widmer2009}. Nevertheless, the prediction accuracy of GNNs becomes relatively low in the fast $\beta$ dynamics compared to that in the slow $\alpha$ relaxation~\cite{Bapst2020,Filion2022}. 
We may therefore conjecture that the dynamics is made difficult to predict owing to the elastic vibrations, which are erased after coarse-graining the short-time dynamics. This decline in the prediction accuracy can be rectified, as shown by a recent study~\cite{Puosi2022}, if we encode short-time vibrations obtained from the actual simulations. However, there are still no methodologies to make predictions solely from a static structure by bypassing the challenges caused by such transient fluctuations.

In this article, we propose a GNN model named ``BOnd TArgeting Network (BOTAN)'' that realizes highly precise predictions of the entire glassy relaxation from a static structure.  BOTAN can learn a target characteristic quantity assigned on edges corresponding to pairs of close-by particles, in addition to the previous target quantity assigned on nodes. BOTAN must be positioned as a straightforward but strong extension of a GNN used in the previous study by Bapst {\it et al.}~\cite{Bapst2020}, which we term as ``node-targeting GNN'' (NT-GNN).
In this study, such a feature of BOTAN is exploited to characterize and learn relative motion; it allows itself to characterize structural rearrangements being unaffected by the elastic fluctuations as well as the local degree of local elastic strains. We show that this modeling allows for highly improved predictions of complex dynamics of slow structural relaxations of glassy systems.

\section{Machine learning model} \label{sec:model}
In this section, we explain the neural network architecture of BOTAN. 
Similarly to NT-GNN, it is based on the interaction network~\cite{Battaglia2016,Battaglia2018} consisting of an encoder--process--decoder architecture, where features of nodes and edges in a graph are mutually computed by exchanging messages between a pair of two-layer multi-layer perceptrons (MLPs) assigned for nodes and edges. This architecture is suitable in extracting intricate relationships from simulation data computed with high precision. Because the network treats node and edge features equivalently, it is possible to regress the network toward the target quantities on edges as well as nodes~\cite{DeZoort2021} by decoding edge embeddings using an MLP. 

\begin{figure}
\centering
\includegraphics[width=0.9\linewidth]{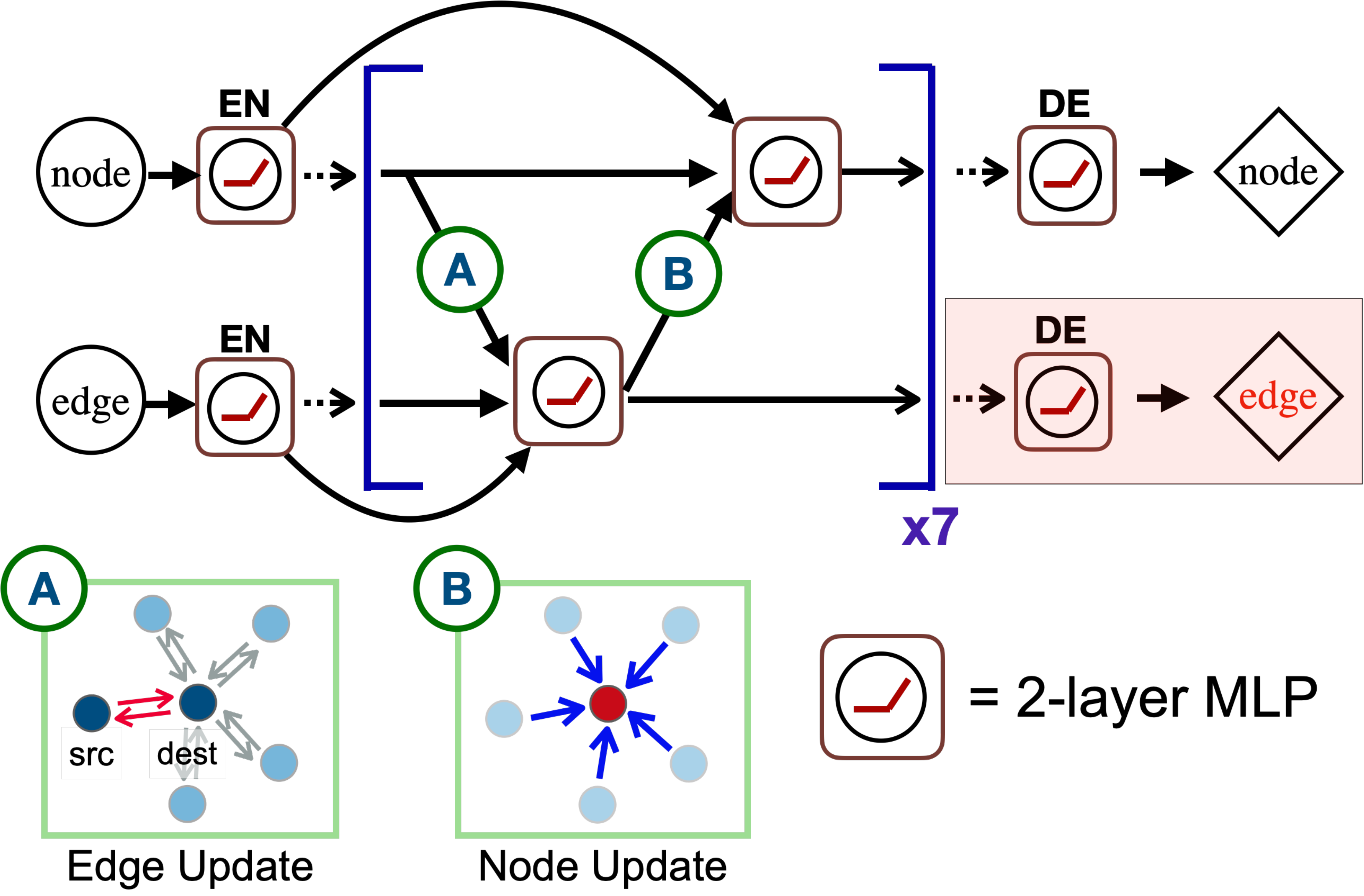}
\caption{Block diagram of the GNN model, BOTAN, indicating the steps of computation in an epoch. ``EN'' and ``DE'' denote ``encoder'' and ``decoder'', respectively. The GNN model has an encoder-process-decoder structure consisting of connected layers of MLPs with two hidden layers of 64 units with ReLU activation. The decoder on the edges is an added part, which does not exist in the NT-GNN.
\label{fig:network} }
\end{figure}

Let $G(V,E)$ be a graph that includes a set of nodes $V$ and edges $E$. For a pair of nodes $v,u \in V$, an edge between $v$ and $u$ is represented as $e_{v,u}$. The neighboring edges of a vertex $v\in V$ are represented as $N(v)$ ($\subset E$). Given input feature vectors $h^{\rm in}_v$ and $h^{\rm in}_e$ for a node $v \in V$ and an edge $e \in E$ respectively, the goal is to compute the corresponding output features $h^{\rm out}_v$ and $h^{\rm out}_e$. As shown in a block diagram in Fig.~\ref{fig:network}, and by denoting an encoder, a multi-layer perceptron (MLP), and a decoder as $\mathrm{EN}(\cdot)$, $\mathrm{MLP}(\cdot)$, and $\mathrm{DE}(\cdot)$,  respectively, each layer used to compute feature vectors for node $v$ is formulated  as follows. First, input feature vectors are encoded by respective encoders 
\begin{eqnarray}
h^0_v = \mathrm{EN}(h^{\text{in}}_v), \label{eq:nodeencoder} \\
h^0_e = \mathrm{EN}(h^{\text{in}}_e). \label{eq:edgeencoder}
\end{eqnarray} 
Then, the update of the edge and node features via MLPs is repeated $n$ times 
\begin{eqnarray}
h^m_{e_{v,u}} &=& \mathrm{MLP}\left(h^{m-1}_{e_{v,u}} \oplus h^{m-1}_v \oplus h^{m-1}_u \oplus h^0_{e_{v,u}}\right) \label{eq:edgeupdate} \\ 
h^m_v &=& \mathrm{MLP} \left(h^{m-1}_v \oplus \sum_{e \in N(v)} h^{m}_{e} \oplus h^{0}_v \right) \label{eq:nodeupdate},
\end{eqnarray}
where $m$ stands for the iteration index of $n$ repeat cycles ($0<m\le n$).
In each cycle, messages are passed between nodes and edges, wherein encoded feature vectors are concatenated together before every update. 
For the results that will be presented in this article, it is critical that edges receive messages from both neighbor nodes $v, u$ in each repeated cycle of message passing in Eq. (\ref{eq:edgeupdate}), as indicated in the inset of Fig.~\ref{fig:network} (``A. Edge update''; an edge is updated bi-directionally, based on the states of nodes on both sides). Furthermore, at each stage of these updates, encoded information, $h_e^0$ or $h_\nu^0$ are concatenated such that the training becomes stable.
Decoders $\mathrm{DE}(\cdot )$, placed as the final layers, yield the output feature on each node and each edge 
\begin{eqnarray}
h^{\text{out}}_v = \mathrm{DE}(h^{n}_v),  \\
h^{\text{out}}_e = \mathrm{DE}(h^{n}_e),  \label{eq:edgedecoder} 
\end{eqnarray}
where, the edge decoder in Eq. (\ref{eq:edgedecoder}) is an additional part of the network that was absent in the NT-GNN. In practice, $\mathrm{EN}(\cdot)$ and $\mathrm{MLP}(\cdot)$ and $\mathrm{DE}(\cdot)$ are all implemented using two hidden layers of 64 units with rectified linear units (ReLU) non-linearity. At every repeat update cycle on edges and nodes in Eqs. (\ref{eq:edgeupdate}) and (\ref{eq:nodeupdate}), information of the neighbor shell (inside the distance of $2.0\sigma_{\rm AA}$) is propagated; therefore, the number of repeat cycles $n$ determines how far the structural information can be incorporated~\cite{Bapst2020,Filion2021}. The prediction accuracy of the model declines as $n$ decreases; this is shown in Appendix~\ref{sec:iteration}. In this study, we set $n=7$, as in the previous study by Bapst {\it et al.}.

\section{Dataset details}  \label{sec:dataset}
In this study, we conduct training and prediction test of BOTAN and NT-GNN by using a dataset of molecular dynamics trajectories of a three-dimensional (3D) Kob--Andersen binary Lennard--Jones mixture~\cite{Kob1995}, wherein the pairwise force is smoothed at its cutoff~\cite{Shimada2018,Dyre2011}.  Interactions between the particles are determined by an interatomic pairwise potential
\begin{equation}
\indent \qquad  u(r) = 4\epsilon_{\mu\nu} \left[ \left( \frac{ \sigma_{\mu\nu}}{r}\right)^{12}
  - \left( \frac{\sigma_{\mu\nu} }{r}\right)^6 \right].
\end{equation}
\noindent
$\mu,\nu \in \{ {\rm A}, {\rm B} \}$ denotes the particle types, where the number of particles belonging to type A (denoted as $N_{\rm A}$) is 80\% of the total number $N=4096$. The interaction energy and particle size are defined as $\epsilon_{\rm AA} = 1.0,\ \epsilon_{\rm AB}=1.5,\ \epsilon_{\rm BB}=0.5,\ \sigma_{\rm AA} = 1.0,\ \sigma_{\rm AB}= 0.8,$ and $\sigma_{\rm BB}=0.88$. In this study, distances, time, and temperature are denoted in units of $\sigma_{\rm AA}$, $\tau = \sqrt{m\sigma_{\rm AA} / \epsilon_{\rm AA}}$, and $\epsilon_{\rm AA}/k_{\rm B}$, with $k_{\rm B}$ the Boltzmann constant
In the remainder of article,  we use the dimensionless units based on the Lennard-Jones potential  and the two species of this mixture are denoted as types A and B.  Trajectories for training and prediction tests are generated as so-called the isoconfigurational ensemble~\cite{Harrowell2004,Widmer-Cooper2006,Harrowell2008,Widmer2009}., wherein 32 simulation runs are conducted with the same particle configuration and different initial velocities. Among the 500 initial configurations, 400 configurations (12,800 runs) are used for training and the other 100 (3,200 runs) are used for testing. 

For improving the accuracy of numerical integration over a large number of steps, the pairwise potential is modified in the track of previous studies as~\cite{Dyre2011,Shimada2018} 
\begin{eqnarray}
U(r) = u(r) - u(r_{\rm c}) - (r-r_c) \left.\frac{du(r)}{dr}\right|_{r=r_{\rm c}}, 
\end{eqnarray}
so that the force and the potential vanish at the cutoff length $r=r_{\rm c}$.  The cutoff lengths are set differently for different combinations of pair species depending on the combination of pair species as $r_{\rm c}=2.5\sigma_{\alpha\beta}$.

\begin{figure}
\centering
\includegraphics[width=0.8\linewidth]{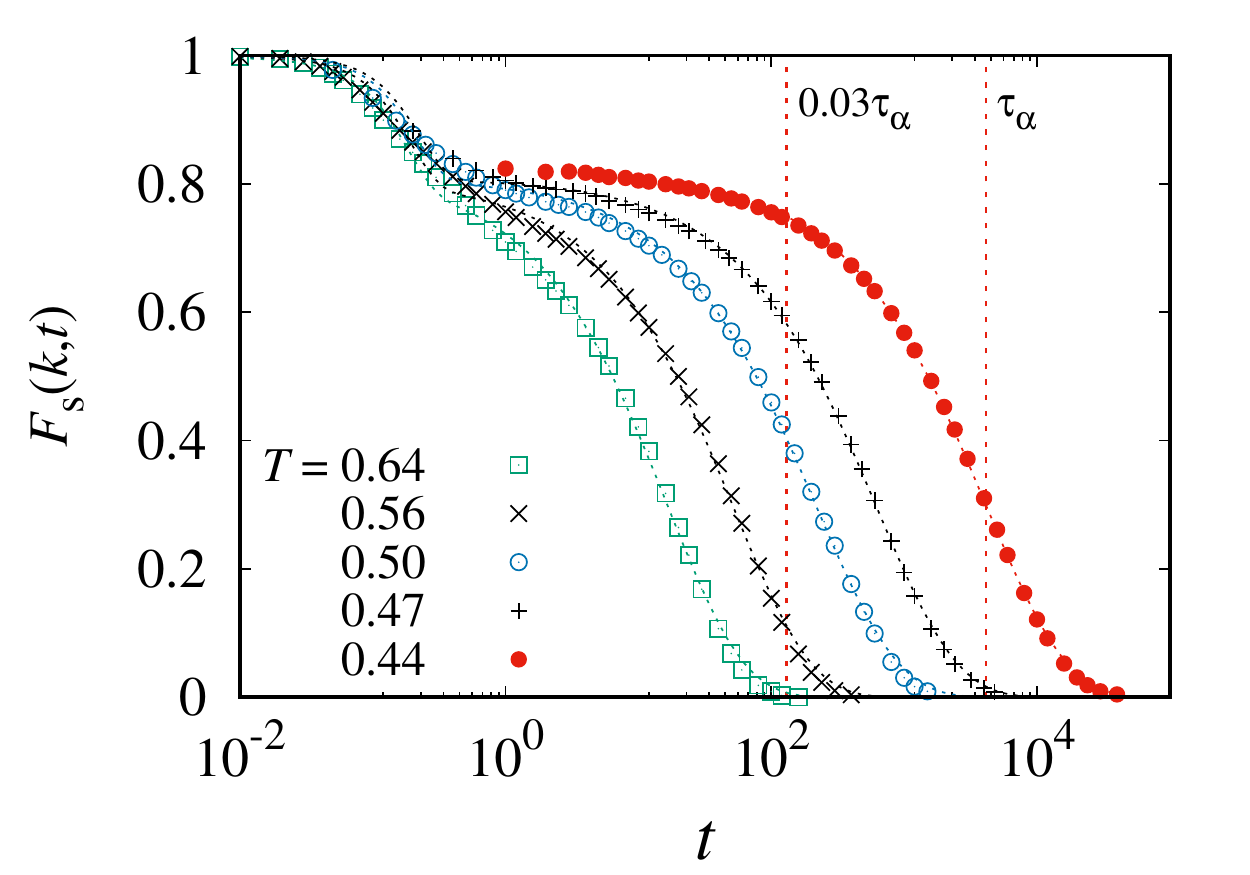}
\caption{The intermediate scattering function $F_{\rm s}(k,t)$ in Eq. (\ref{eq:fskt}) is shown for temperatures $T=0.44,$ 0.47, 0.50, 0.56, and 0.64 as functions of time. Vertical dotted lines indicate time points corresponding to $0.03\tau_\alpha$ and $\tau_\alpha$ for $T=0.44$, which are mainly investigated in this article. \label{fig:fskt} }
\end{figure}

The procedure of dataset generation is similar to that in the previous study~\cite{Bapst2020}, except that the system size is fixed. For all simulations, the number density of the system is fixed at $N/V = 1.2$. The time step of numerical integration is kept at $\Delta t=10^{-3}$ throughout. We set three target temperatures:  $T=0.44,\ 0.50,$ and 0.64. 
In Fig. \ref{fig:fskt}, the self-parts of the intermediate scattering functions for particles of type A
\begin{equation}
    F_{\rm s} (k,t) = \frac{1}{N_{\rm A}}\sum_{j\in {\rm A}}\exp \left\{ i\bm{k}\cdot [ \bm{r}_j (t+t_0) - \bm{r}_j(t_0) ] \right\},  \label{eq:fskt}
\end{equation}
are shown for these temperatures (and additionally $T=0.47$ and 0.56),
where the wavenumber $k$ is set at $2\pi / \sigma_{\rm AA}$. The  $\alpha$-relaxation times $\tau_\alpha$ are evaluated as 4120,\ 217,\ and 16.5,
for $T=0.44,\ 0.50,$ and 0.64, respectively, by fitting it to a double stretched exponential function~\cite{Sengupta2013,Shiba2019}.  In the Letter, the time points $t = 130\ (0.03\tau_\alpha)$ and $4120\ (\tau_\alpha )$ are mainly investigated at the lowest temperature $T=0.44$,  which is also indicated in Fig. \ref{fig:fskt} with vertical dotted lines. 

For each of these three target temperatures, 500 independent particle configurations are generated as follows. After keeping the temperature at $T=5.0$ for the time lapse of $t_{h} = 10^4$, the system is cooled rapidly (in $t_{\rm cool} = 20$) to the target temperature. The temperature is subsequently kept constant by using the Nos\'e-Hoover thermostat until the system reaches the steady state up to the time scale of 40$\tau_\alpha$. The final particle configuration thus obtained is used as the initial configuration at $t=0$ in the production run for generating the dataset.

From the 500 initial configurations, we introduce the isoconfigurational ensemble~\cite{Harrowell2004,Harrowell2008} by running 32 separate microcanonical ($NVE$) simulations all starting from the same configuration; however, the initial velocities are randomly given from the Maxwell-Boltzmann distribution at the target temperature. These simulations are conducted approximately up to $30\tau_\alpha$, where the particle configurations are sampled at logarithmically increasing time points, as summarized in Table \ref{tab:my_label}. Among the 500 initial configurations, 400 configurations (12,800 runs) are used for training and the other 100 (3,200 runs) are used for testing.

\begin{table*}
\centering
    \begin{tabular*}{15cm}{@{\extracolsep{\fill}}|r||l|l|l|l|l|l|l|l|l|l|}
        \hline
        {tagging} & {\it 1} & {\it 2}  &  {\it 3} &  {\it 4} &  {\it 5} &  {\it 6} & {\it 7}  & {\it 8} & {\it 9} & {\it 10} \\
         \hline \hline
         $T=0.44$ & 0.13 & 1.30  &  13.0 &  {\bf 130} &  412 &  1300$\ $ & {\bf 4120}$\ $ & 13000$\ $  & 41200$\ $  & 130000$\ $ \\
        \hline
         $0.50$ & 0.108$\ $ & 0.434 & 2.17 &  6.86 & 21.7 &  68.6 & 217  &  686  &   2170 &  6860  \\
         \hline
         $0.64$ &  & 0.100 & 0.165 & 0.52 & 1.65 &  5.2 &  16.5 & 52   &  165 &  520  \\ 
         \hline
    \end{tabular*}
    \caption{Time points used for generating the training data (in the nondimensional LJ units). At each temperature, we set 9 or 10 time points for which isoconfigurational ensemble of 32 trajectories are taken. 
    \label{tab:my_label}}
\end{table*}

\section{Procedures of machine learning}

In this study, we conduct training and prediction test of BOTAN and NT-GNN by using a dataset of molecular dynamics trajectories of a three-dimensional (3D) Kob--Andersen binary Lennard--Jones mixture~\cite{Kob1995}, wherein the pairwise force is smoothed at its cutoff~\cite{Shimada2018,Dyre2011}.  In the remainder of article,  we use the dimensionless units based on the Lennard-Jones potential  and the two species of this mixture are denoted as types A and B.  Trajectories for training and prediction tests are generated as so-called the isoconfigurational ensemble~\cite{Harrowell2004,Widmer-Cooper2006,Harrowell2008,Widmer2009}., wherein 32 simulation runs are conducted with the same particle configuration and different initial velocities. 
Among the 500 initial configurations, 400 configurations (12,800 runs) are used for training and the other 100 (3,200 runs) are used for testing. The relevant details are given in Appendix~\ref{sec:dataset}.

A graph for training or prediction test is constructed based on the particle configuration wherein the particles are represented by its nodes. Edges are assigned to pairs of particles $i$ and $j$ within a distance threshold value $r_{ij}(t=0) \le r_e$ so that the neighborhood relation can be represented. We choose the threshold length $r_e = 2.0\sigma_{\rm AA}$ that has the best predictive performance regarding particle propensity~\cite{Bapst2020} ($\sigma_{\rm AA}$ denotes interaction radius between a pair of particles of type A). 

The constructed graphs are then fed into the model by encoding particle types (A or B) into the nodes and 3D relative positions between particles ($\bm{r}_{ij} = \bm{r}_i - \bm{r}_j$) into the edges. 
Specifically, for input feature vectors $h^{\text{in}}_v \in \{A, B\}$ and $h^{\text{in}}_e$ ($=\bm{r}_{ij}$) $\in \mathbb{R}$, the first layers of the node and edge encoders (Eqs.~\eqref{eq:nodeencoder} and \eqref{eq:edgeencoder} are implemented by linear transformations $f_{\rm n}: \mathbb{R} \to \mathbb{R}^d$ and
$f_{\rm e}: \mathbb{R}^3 \to \mathbb{R}^d$, respectively, where $d$ is the dimension of the hidden layers (64 in our case). For the node feature, type A and B are converted into real numbers $0$ and $1$ beforehand, and both the input features on nodes and edges are treated as real number with 32-bit single-floating point precision.

In NT-GNN~\cite{Bapst2020,Filion2022},  the node output feature $h^{\text{out}}_v$ was exploited for learning and prediction test, whereas the edge output feature $h^{\text{out}}_e$ is additionally extracted for these tasks in BOTAN. 
In this study, the target quantities for training and prediction tests are the particle displacement (traveling distance) of each particle $i$ 
\begin{equation}
\mathcal{S}_i(t) = |\bm{r}_i (t) - \bm{r}_i(0) |
\end{equation} 
and pair-distance changes characterizing {\it relative motion} between neighbor pairs 
\begin{equation}
\mathcal{E}_{ij}(t) = r_{ij}(t) - r_{ij}(0), \label{eq:pairdistancechange}
\end{equation}
respectively.  Both of these target quantities of learning are averaged over isoconfigurational ensembles, for which explicit notations are abbreviated. The GNN models are independently trained and tested at different timepoints.

The learning setups and hyperparameters in the present study are similar to those used by Bapst {\it et al.}~\cite{Bapst2020}. The training dataset is learned to minimize the loss function, wherein the neural network is optimized using the Adam algorithm without norm regularization, with a learning rate of $10^{-4}$ and a standard implementation in PyTorch.
The data loaded for training are augmented by applying random rotations of the simulation box to the particle positions, as all the target quantities  for learning are scalar variables that are rotationally invariant.

The loss functions for learning are defined by 
\begin{equation}
\mathcal{L}_{\mathcal{M}} = p \mathcal{L}_{\mathcal{S}} + (1-p)\mathcal{L}_{\mathcal{E}},
\end{equation}
where $\mathcal{L}_{S}$ and $\mathcal{L}_{E}$ are standard $\mathcal{L}_2$-norm loss functions respectively defined for deviations of particle propensity $\mathcal{S}_i(t)$ and pair-distance changes $\mathcal{E}_{ij}(t)$. These may be written as 
\begin{eqnarray}
\mathcal{L}_\mathcal{S} &=& \sum_{j=1}^N \left[ \hat{\mathcal{S}}_j(t;\{  \bm{r}_i\} ) - \langle {\mathcal{S}}_j(t)\rangle_{\rm IC} \right]^2 \\   
   \mathcal{L}_\mathcal{E} &=& \sum_{j=1}^N \sum_{k\in {\rm n.n.}} \left[ \hat{\mathcal{E}}_{jk}(t;\{ \bm{r}_i\}) - \langle {\mathcal{E}}_{jk}(t) \rangle_{\rm IC} \right]^2, 
\end{eqnarray}
where $\langle \cdot \rangle_{\rm IC}$ represents the average over the isoconfigurational ensemble, and $\hat{\mathcal{S}}_j(t,\{ \bm{r}_j \})$ and $\hat{\mathcal{E}}_{jk}(t,\{ \bm{r}_i \})$ denote the values of the particle propensity and pair-distance change as the output from the GNN. Using loss function $\mathcal{L}_{M}$, the model can be trained via both $\mathcal{S}_j(t)$ and $\mathcal{E}_{jk}(t)$ simultaneously, and further, these quantities can be inferred simultaneously as well. Parameter $p$ is treated as a hyperparameter $(0\le p\le 1)$, as it decides the weight of learning between nodes and edges. Two limiting cases are $p=0$, wherein the model only learns pair-distance changes $\mathcal{E}_{ij}(t)$, and $p=1$, which reduces to NT-GNN~\cite{Bapst2020}. For simultaneous learning on nodes and edges ($0<p<1$), we set $p=0.4$ in this study. This choice is justified by the ablation experiment described in Appendix \ref{sec:anticorr}.

\begin{figure}
\includegraphics[width=0.9\linewidth]{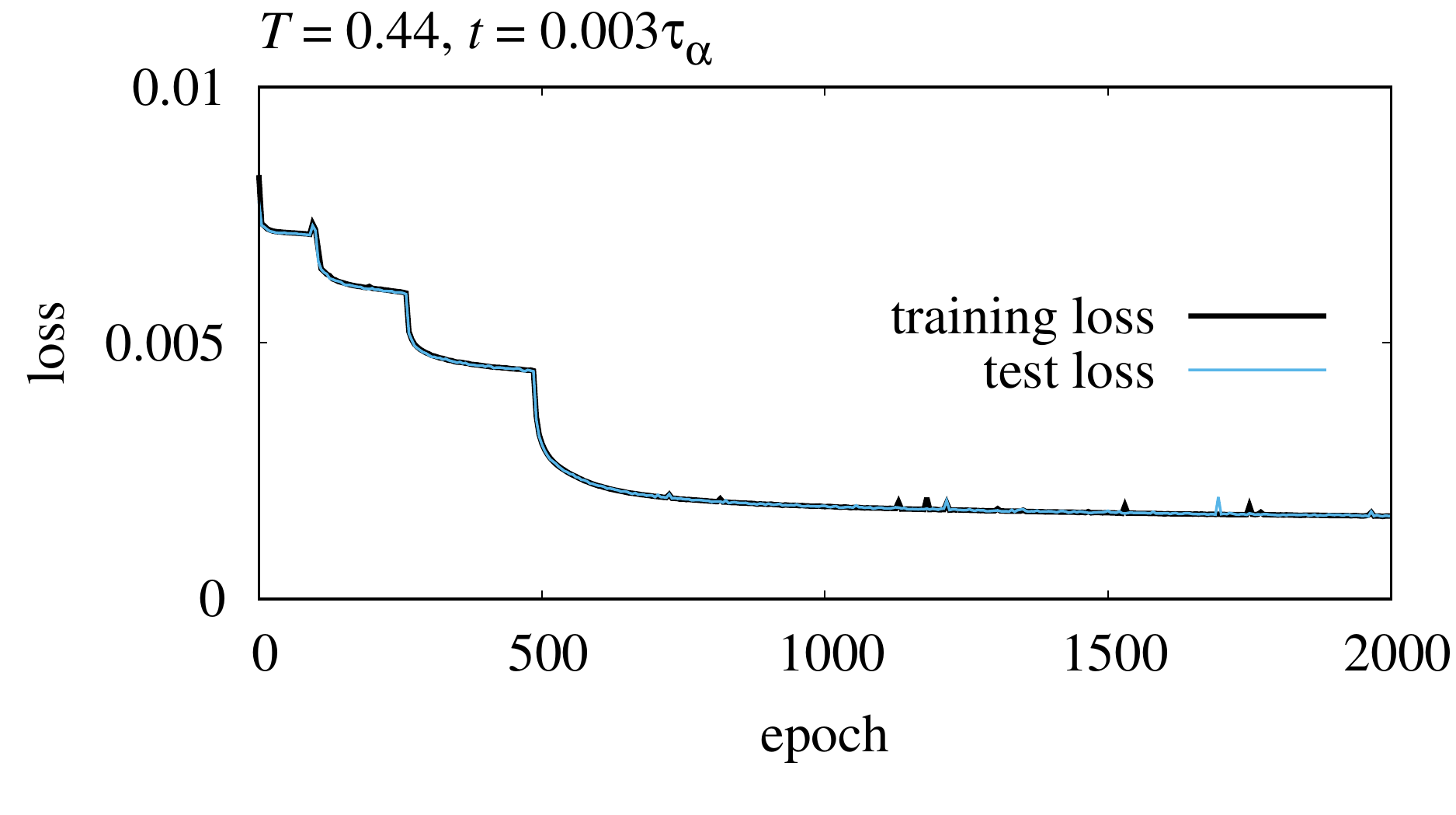} 
\caption{\label{fig:pretrain}
Training and test losses (left axis) for $t=13\ (0.003\tau_{\alpha})$ at $T=0.44$ for the pretraining of pair-distance changes $\mathcal{E}_{ij}(t)$ as an edge feature, with $p=0$. 
}
\end{figure}

In all these training scenarios, all the particles (all the particle pairs) are learned together, regardless of the particle types. No minibatching of graph inputs are applied in the training, except that the minibatch size of the graph input is fixed at 5 in the training of NT-GNN ($p=1$). Although Bapst {\it et al.} applied no minibatching in their previous study~\cite{Bapst2020}, this difference does not significantly change the results presented in this article. The training with NT-GNN required approximately 2 hours when using one NVIDIA A100 Tensor Core GPU (40 GB SXM). Essentially the same amount of time is required for training BOTAN, except that there may exist a substantial overhead arising from computation of interparticle distances, for which the actual elapsed time depends on the implementation. 

In fact, many training epochs are required if we start the training of BOTAN with random weight parameters. The reason is that, as will be discussed in Sec.~\ref{sec:edge}, there is a simple relation between the initial distance $r_{ij}(t=0)$ and its change after time $\mathcal{E}_{ij}(t)$ when the interval time $t$ is short. 
Such a relation defines a ``metabasin'' to which many weight parameters are largely optimized. Therefore, in this study, we start all the training of BOTAN $(p<1)$ with weight parameters that are adjusted by pretraining on edges at a low temperature $T=0.44$.  This pretraining is performed for 2,000 epochs with the pair-distance changes, by keeping $p=0$ and using trajectory data at a short time $t=13$, which results in a sufficient convergence of the loss functions, as shown in Fig. \ref{fig:pretrain}. 
Such pretraining allows the main part of the training scenario to be efficient enough to finish within 1,000 epochs.

\section{Results}
\subsection{Predicting pair-distance changes as an edge feature} \label{sec:edge} 

\begin{figure}
\centering
\includegraphics[width=\linewidth]{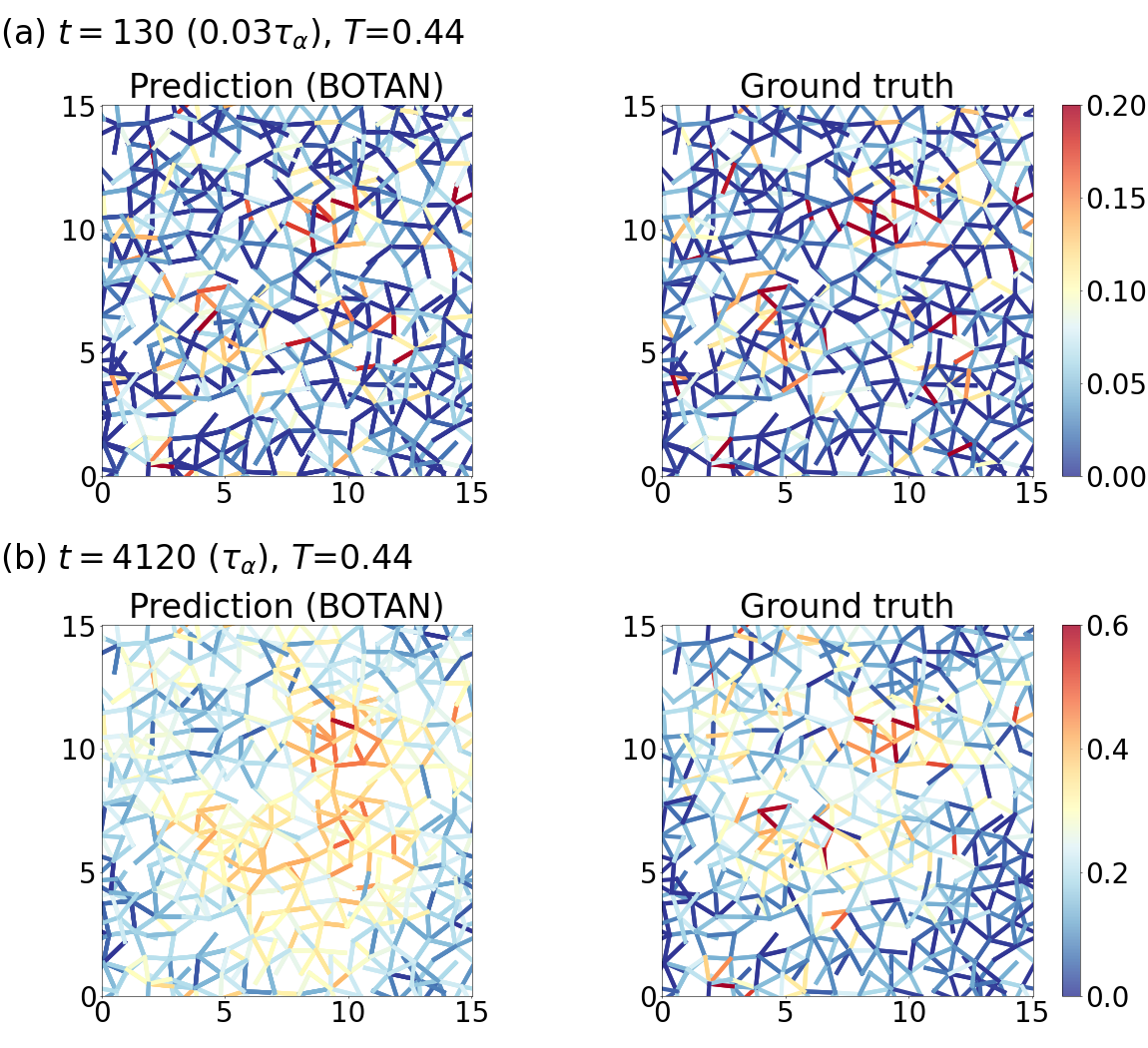} \\
\caption{\label{fig:edge}
Distribution of the predicted and ``actual'' (ground truth) pair distance change $\mathcal{E}_{ij}(t)$ at temperature $T=0.44$ for  $t=$ (a) 130 (0.03$\tau_\alpha$) and (b) 4120 ($\tau_\alpha$), respectively, which are plotted in color maps  projected on each segment. Segments represent pairs within a distance of $r< 1.35$. A specific cross section ($11.1< z< 11.9$) is cut out of the 3D box system.
}
\end{figure}

\begin{figure}
\centering
\includegraphics[width=0.75\linewidth]{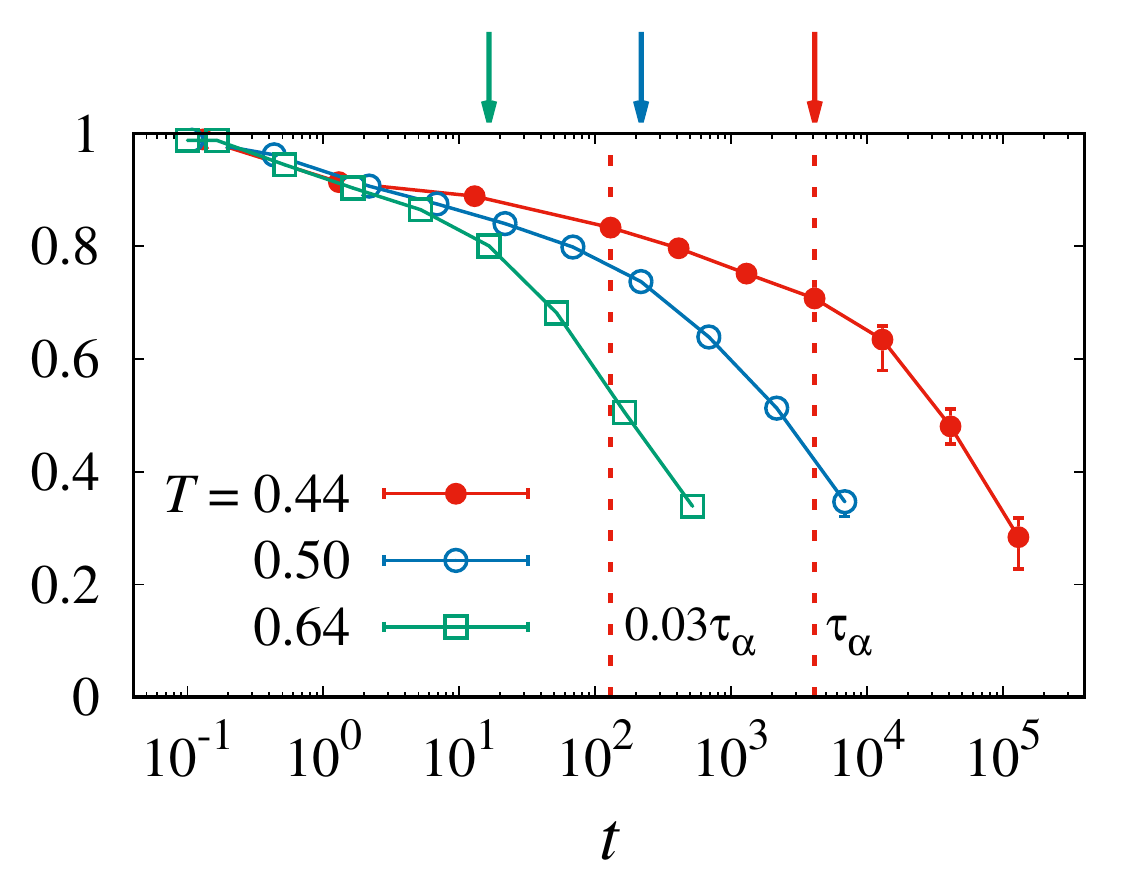} \\
\caption{\label{fig:pearson_edge}
Pearson correlation coefficients between predicted and ground truth values of $\mathcal{E}_{ij}(t)$ plotted as a function of time $t$ at $T=0.44$, $0.50$, and $0.64$. Error bars depict the median, best, and worst of five independently trained models. The two dotted lines indicate $0.03\tau_\alpha$ and $\tau_\alpha$ for $T=0.44$, and the arrows show the $\alpha$-relaxation times for respective temperatures.}
\end{figure}

First, we investigate the predictive ability of BOTAN when it is trained only with pair-disntance change $\mathcal{E}_{ij}(t)$, namely, the case of $p=0$. 
In Fig.~\ref{fig:edge}, prediction results of the pair-distance change $\mathcal{E}_{ij}(t)$, obtained for $t=$ (a) 130 and (b) 4120 using BOTAN trained at respective time points and at the lowest temperature $T=0.44$, are shown for a particle configuration in the test dataset. The data are shown in color maps projected onto segments satisfying $r_{ij}(t=0) < 1.35$ in a two-dimensional cross section. The ``actual'' pair-distance changes (the isoconfigurational ensemble average) from the simulation are also shown for each as the ground truth. The time points correspond to $0.03\tau_\alpha$ and $\tau_\alpha$, where $\tau_\alpha$ is the $\alpha$-relaxation time.   
BOTAN effectively discerns which specific pairs become separated, particularly for the short time in (a). 
Over a long time, as in (b), the prediction deviates from the ground truth, but it still captures pair-level propensity for distance changes as well as the spatial contrast between mobile and immobile regions. In Fig.~\ref{fig:pearson_edge}, the prediction accuracy of BOTAN over $\mathcal{E}_{ij}(t)$ is quantified by using the Pearson correlation coefficient~\cite{Bapst2020,Filion2021,Filion2022}
\begin{equation} 
\rho = \frac{\sum_i (x_i-\overline{x})(y_i-\overline{y})}{ \sqrt{\sum_i (x_i-\overline{x})^2} \sqrt{\sum_i (y_i-\overline{y})^2}},
\end{equation}
which expresses the proximity between the predicted ($\{ y_i\}$) and ground truth ($\{ x_i\}$) data. Here, $\overline{x}$ and $\overline{y}$ denote the average of $x_i$ and $y_i$.  This coefficient should be quantified carefully to ensure that the comparison is made over equivalent particle pairs. Therefore, among pairs of type A particles (pairs related to type B are excluded), pairs within the first neighboring distance $r_{ij}(t=0) < 1.35$ are chosen in this comparison, because neighbor pairs in the first and second neighbor shells are expected to exhibit different distance changes on average.  The prediction accuracy thus evaluated is high in the short time, and {\it monotonically} declines with time. 

\begin{figure}
\centering
\includegraphics[width=0.9\linewidth]{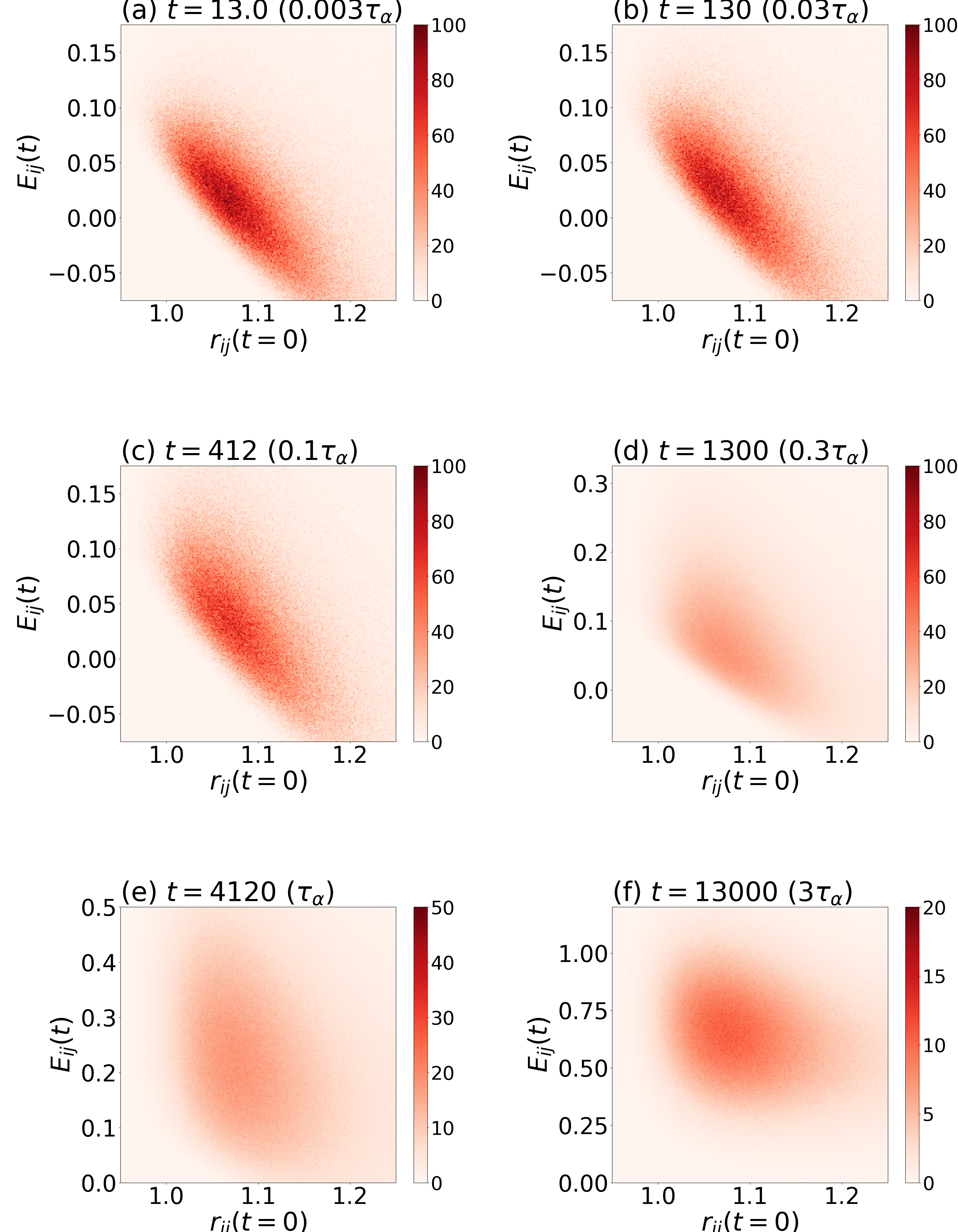}
\caption{The probability distribution function showing the relationship between the initial pair distance $r_{ij}(t=0)$ and pair-distance changes $\mathcal{E}_{ij}(t)$ 
in the simulation data used for training at $T=0.44$, for time points $t=$(a) 13.0, (b) 130,  (c) 412, (d) 1300, (e) 4120, and (f) 13000.   
\label{fig:pdf_hist} }
\end{figure}

The high prediction accuracy in the short-time region originates from a simple relation between the pair-distance change $\mathcal{E}_{ij}(t)$ and initial distance $r_{ij}(t=0)$. Fig.~\ref{fig:pdf_hist} shows a probability distribution function map as functions of these quantities for the closest neighbor pairs of type A particles in the training dataset at time points ranging from $t=13$ to $13000$ at $T=0.44$.  In the short-time region ($t< 0.1\tau_\alpha$ = 412), a simple anti-correlation exists, wherein the pair-distance change $\mathcal{E}_{ij}(t)$ remains one order of magnitude smaller than the particle radii for most of the neighbor pairs. This quantity, therefore, characterizes strains induced by the elastic vibrations. The anti-correlation gradually disappears but persists up to $t=\tau_\alpha$ as the time increases, making the relation between the two less trivial. The anti-correlation determines the metabasin in the landscape of the loss function in the vast weight-parameter space. The number of training epochs required to optimize BOTAN drastically increases as the anti-correlation gradually disappears for a longer time $t$, as the loss landscape becomes more complex to make the metabasin less ``noticeable''. Therefore, we pretrained BOTAN over the pair-distance changes at $T=0.44$ after $t=13$, then we used the weight parameters for training with varying time. As the metabasins in the parameter space are likely to be located close to each other between the different time points, the usage of pretrained models would allow us to concentrate on optimizing the weight parameters around the metabasins and thus reduce the number of training epochs required for the main part of training.

\subsection{Predicting particle propensity for motion as a node feature} \label{sec:nodeedge} 
\begin{figure*}
\centering
\includegraphics[width=0.9\linewidth]{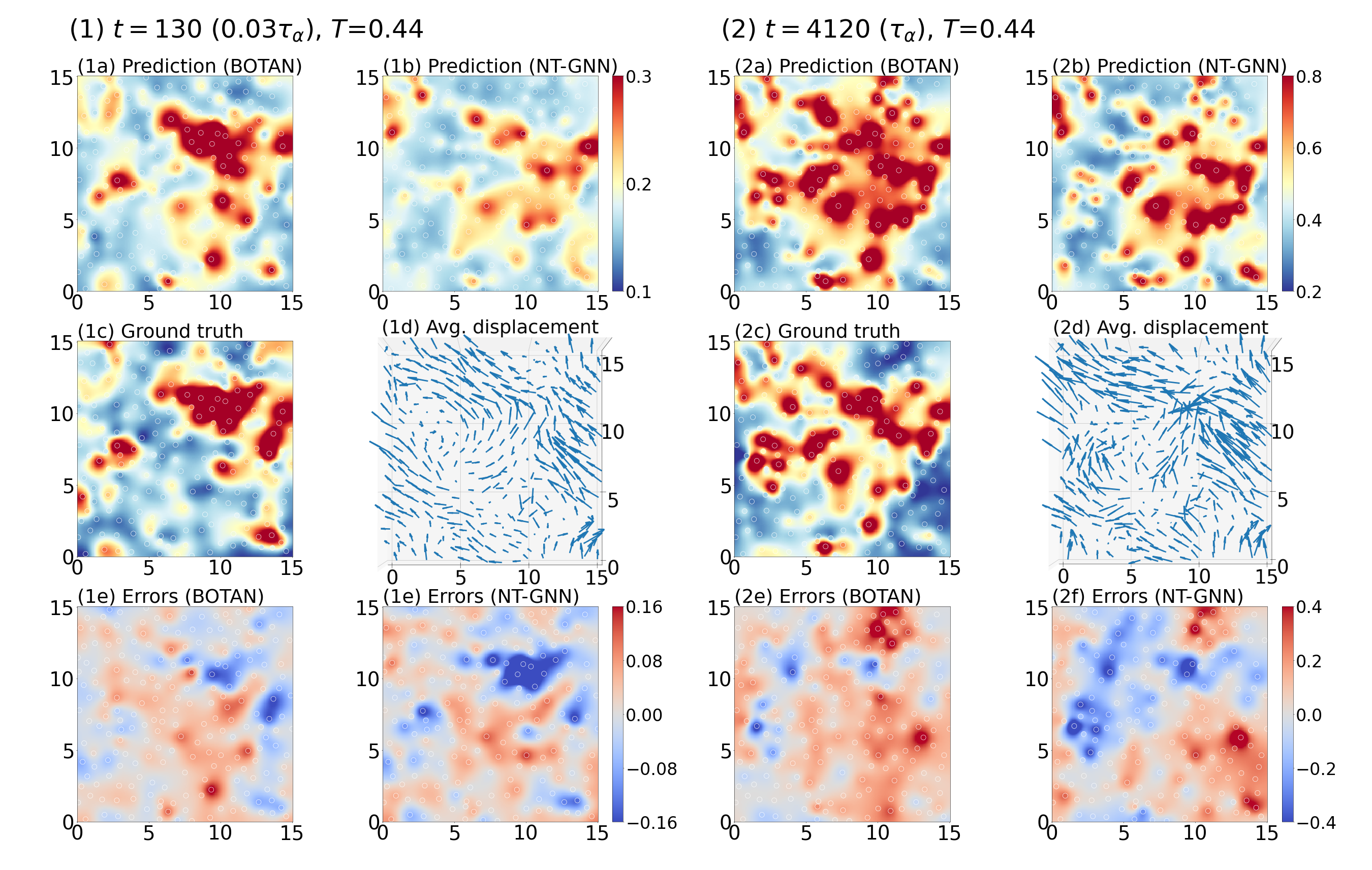} 
\caption{\label{fig:nodeedge} 
Comparison of predictions and ground truth data for $t=$ (1) 130 ($0.03\tau_\alpha$) and (2) 4120 ($\tau_\alpha$), both at the lowest temperature $T=0.44$: 
(a) and (b) Particle propensity $\mathcal{S}_i(t)$ predicted using (a) BOTAN and (b) NT-GNN, respectively.  Color maps are linearly interpolated between particles. (c) Distribution of $\mathcal{S}_i(t)$ as the ground truth data.
(d) Local displacement field averaged over neighbors. (e) and (f) Error map for predictions of (e) BOTAN and (f) NT-GNN, corresponding to (a) and (b), respectively. All data are plotted for the same 3D particle configuration, in the cross section $11.1<z<11.9$, except (e), in which $10.9 < z < 12.1$ is shown.
}
\end{figure*}

Because BOTAN has a pair of decoders that respectively computes output features for nodes and edges, it 
is possible to train BOTAN simultaneously with target quantities for both, particle propensity $\mathcal{S}_i(t)$
and pair-distance changes $\mathcal{E}_{ij}(t)$.  This is realized by setting the loss function as a weighted sum of $\mathcal{L}_2$-norm losses for  $\mathcal{S}_i(t)$ and $\mathcal{E}_{ij}(t)$. The model is then trained with other hyperparameters remaining unchanged.  

We compare the prediction ability of BOTAN with that of NT-GNN by Bapst {\it et al}. Fig. \ref{fig:nodeedge} shows the prediction results for (1) $t=130$ ($0.03\tau_\alpha$) and (2) $t=4120$ ($\tau_\alpha$) at $T=0.44$ of (a) NT-GNN and (b) BOTAN, in addition to (c) the ``actual'' propensity directly evaluated from the trajectory data in the test dataset. As all the snapshots are shown in the same cross section as in Fig. \ref{fig:edge} (a), the spatial correspondence can be seen between the predictions of $\mathcal{S}_i(t)$ and $\mathcal{E}_{ij}(t)$.  
BOTAN clearly makes better predictions of spatially heterogeneous patterns of particle propensity than NT-GNN. 
In (d), 3D vector plots are also provided to show the displacement field locally averaged over nearest neighbors, defined by 
\begin{equation}
\langle \Delta \bm{r}^{\rm ave}_i(t) \rangle_{\rm IC}
= \left\langle \frac{\sum_j \left[ \bm{r}_j (t) - \bm{r}_j(0) \right]\ \Theta \left( r_{f} - r_{ij}(0) \right) }{ \sum_j  \Theta \left( r_{f} - r_{ij}(0) \right)  } \right\rangle_{\rm IC}.
\label{eq:cgdr}
\end{equation}

where $\Theta (x)$ denotes Heaviside’s step function and $r_f = 1.35$ is the cutoff length of coarse-graining.  Each is plotted in a cross-section with a thickness of 1.2, where the 3D vectors are enlarged for better visibility. These plots can extract  particle motion taking place as a consequence of quasi-localized and phonon vibrations~\cite{Mizuno2017,Manning2021,Tanaka2022} after removing the effect of interparticle rearrangement.  For both $t=130$ and 4120, we observe collective motion on the length scale considerably exceeding the particle size, whereas the flows are more aligned in the former case, suggesting the effect of phononic motion.
In (e) and (f), error maps of predictions are shown for BOTAN and NT-GNN, respectively, showing the existence of spatial heterogeneity in prediction errors. By comparing the error map of NT-GNN (f) with the locally-averaged displacement field in (d), we find that these spatial patterns are corresponding -- that is, the locally-averaged displacements tend to be parallel with the contour lines of the prediction error map, especially in the region where the displacement vectors are aligned. This clear spatial correlation suggests that the collective fluctuations, which are less intimately related to the local structure around each particle, are a possible reason for the decline in prediction accuracy of NT-GNN~\cite{Bapst2020,Filion2022}.

We next assess the prediction accuracy of BOTAN in comparison with NT-GNN. Because particles of types A and B are expected to exhibit different diffusivity over a long period, the prediction accuracy is evaluated in terms of the Pearson correlation coefficient using only the data for type A particles.  Figure~\ref{fig:pearson} shows how accurately BOTAN and the NT-GNN predict the particle propensity $\mathcal{S}_i(t)$ for three different temperatures, $T=$ (a) 0.44, (b) 0.50, and (c) 0.64. The trend of time dependence for NT-GNN is in agreement with the previous results~\cite{Bapst2020}: in the shorter time before reaching the plateau region ($t < 0.2$), the Pearson correlation coefficient assumes high values, then falls below 0.5, and afterwards gradually increases to reach its peak at around the $\alpha$-relaxation time. Conversely, BOTAN outperforms NT-GNN in its predictive accuracy over the entire temporal range and for all temperatures under investigation. 
 
To this end, we note that the pair-distance change in Eq. (\ref{eq:pairdistancechange}) has its own significance despite the extreme simplicity of the definition. Because the pair-distance changes characterize relative motion, they can capture structural {\it rearrangements} .
This feature is distinct from standard quantities defined on the basis of particle displacements that is affected by non-local phononic fluctuations, and thus enables the model to better distinguish the rearranging hotspot separating non-local effects. 
Similar ideas have also been implemented in the method of counting the replacement of neighboring pairs~\cite{Harrowell2008,Yunker2009,Schall2012}, also termed as ``bond breakage''~\cite{Yamamoto1998b,Shiba2012,Guiselin2022}.  Furthermore, the pair-distance changes can characterize the extent to which each pair tends to change owing to the {\it local shear deformation} around rearranging cores; this aspect of structural changes has long been addressed using the concept of ``shear transformation zones''~\cite{Falk1998,2007Ogata,Manning2021,Mandaptu2022}.  By learning how  strains of non-rearranging regions due are distributed  and how rearranging hot spots are localized, BOTAN autonomously fixes the errors in the prediction of particle displacements, and overall acquires unprecedented predictive ability regarding the glassy dynamics.

\subsection{Comparison with prediction accuracy of other quantities by NT-GNN} \label{sec:falk} 
Finally, we try to compare the level of prediction accuracy   that can be achieved  by training NT-GNN by replacing target variables into a quantity in which the effect of collective motion is restricted to a certain extent. For this purpose, we introduce and briefly examine two quantities that characterize particle rearrangements being less affected by such collective motion. 
One is the relative displacement that they would have under an uniform strain 
\begin{eqnarray}
\mathcal{R}_i(t) = \frac{1}{N_{\rm n.n.}}\! \left| \sum_{j\in {\rm n.n.}} \left\{  [ {r}_i^\alpha (t) -{r}_j^\alpha (t) ]  - [ {r}_i^\beta (0) -  {r}_j^\beta (0) ]  \Lambda_{\alpha\beta} \right\} \right| \quad  \nonumber \label{eq:relative} \\
\Lambda_{\alpha\beta} = \delta_{\alpha\beta} + \epsilon_{\alpha\beta}, \nonumber
\end{eqnarray}
wherein the vector and tensor components $\alpha,\beta \in \{ x,y,z\}$ are explicitly denoted for clarity. Here,  $\epsilon_{\alpha\beta}$ represents the local strain tensor, which is evaluated from change in the local arrangement of neighboring particles in the distance of 1.6, and $N_{\rm n.n.}$ denotes the number of nearest neighbors around $i$-th particle.  When this local strain reduces to zero ($\epsilon_{\alpha\beta}=0)$, which is valid in the limit of $t\rightarrow 0$, this quantity reduces to a more easily interpretable form $\mathcal{C}_i(t) = \left| \Delta \bm{r}_i(t) - N_{\rm n.n.}^{-1}\sum_{j\in {\rm n.n.}}  \Delta \bm{r}_j(t)\right| $,  clearly manifesting itself as the relative displacement with respect to neighbor  environments~\cite{Shiba2016,Shiba2019}. 

The other quantity is a further variant but widely used ~\cite{Falk1998,2007Ogata,Peng2011,Schall2012,Schoenholtz2014}, and
defined as the mean-square of difference between the actual displacement and the uniform strain displacement of the neighbors
\begin{equation}
\mathcal{D}_i^2(t) = \frac{1}{N_{\rm n.n.}} \sum_{j\in {\rm n.n.}} \sum_\alpha \left\{ [ {r}_i^\alpha (t) - {r}_j^\alpha (t) ]  - [{r}_i^\beta (0) -{r}_j^\beta (0) ]  \Lambda_{\alpha\beta}  \right\}^2,  \nonumber \label{eq:falk}
\end{equation}
where we employ its square root $\mathcal{D}_i(t)$ as the target of learning. This quantity is close to  $\mathcal{R}_i(t)$ in its form; the difference lies in the squared sum that is more distinctly affected by a specific $i$-$j$ pair if a specific $i$-$j$ pair becomes further away than other pairs. Therefore,  $\mathcal{D}_i(t)$ is more susceptible to hot spots where particle rearrangement preferably occurs. 

\begin{figure}
\centering
\includegraphics[width=0.9\linewidth]{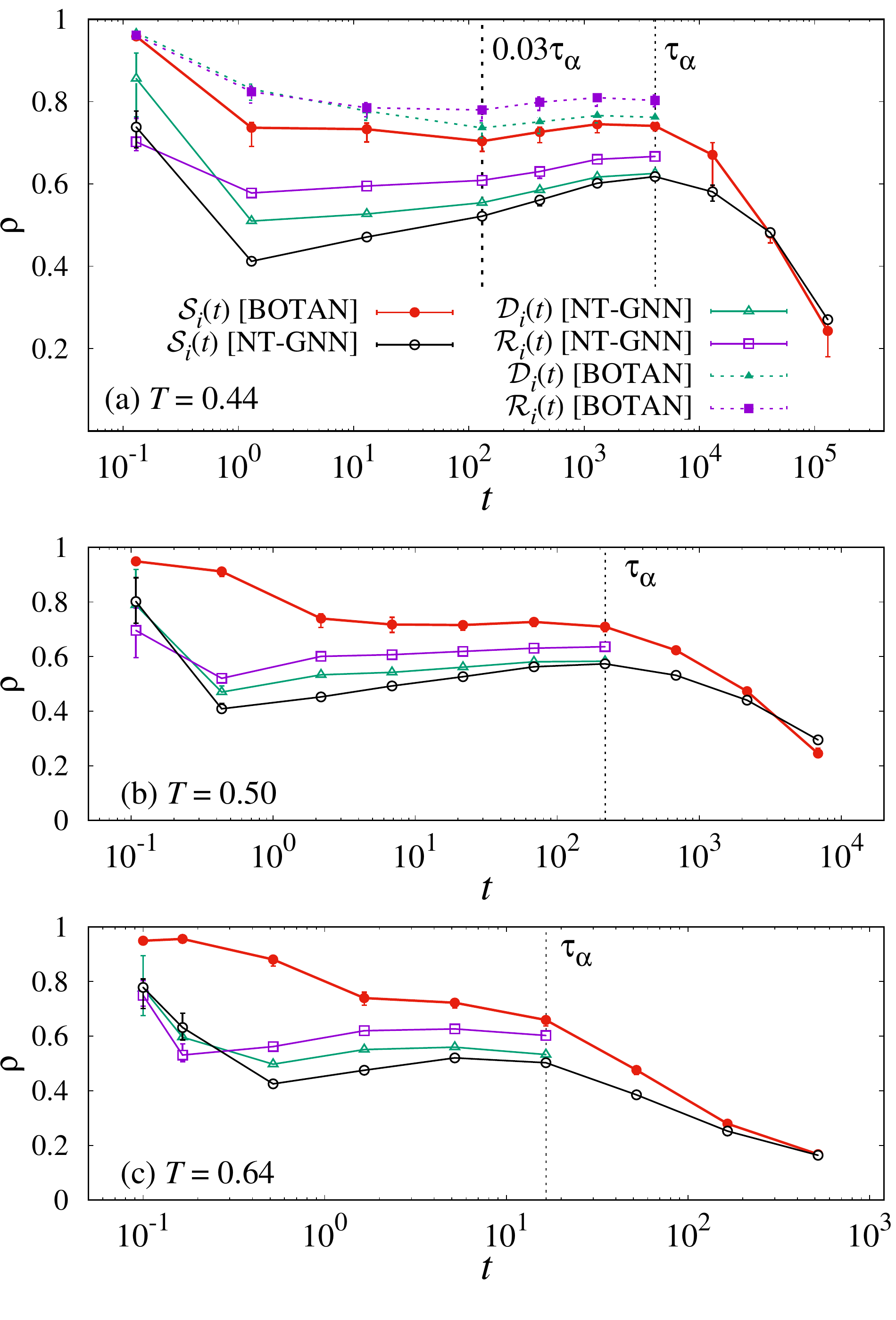}
\caption{\label{fig:pearson}
Pearson correlation coefficients $\rho$ between predicted and actual particle propensities $\mathcal{S}_i(t)$, representing the prediction accuracy of BOTAN and NT-GNN, are shown as functions of time $t$ for $T=$ (a) 0.44, (b) 0.50, and (c) 0.64. Additionally, those of $\mathcal{R}_i(t)$ and $\mathcal{D}_i(t)$ obtained by NT-GNN are also shown for comparison. Error bars show the median, best, and worst of five independently trained models.
}

\end{figure}

In Fig.~\ref{fig:pearson}, Pearson correlation coefficients between predictions of NT-GNN and the actual values of neighbor-relative particle propensities, $\mathcal{R}_i(t)$ and $\mathcal{D}_i(t)$, are shown and further, in (a), this correlation coefficient is also plotted for the prediction made by BOTAN.  The correlation coefficient is evaluated only for type-A particles, as a proximity between the predicted and ground truth values. The prediction accuracy of NT-GNN improves over these quantities, especially in the time regions shorter than $0.1\tau_\alpha$. This clearly implies that spatially extended static fluctuations cause decline in the prediction accuracy of $\mathcal{S}_i(t)$ by NT-GNN in the short time, and that this decline in the prediction accuracy is recovered because, in these quantities, effect of these spatially-extended static fluctuations are removed to a certain extent.

\section{Conclusion}
In conclusion, we have introduced a GNN model that realizes faithful predictions of glassy dynamics over the entire temporal range, with accuracy greatly improved from previous models, including the GNN model by Bapst {\it et al}~\cite{Bapst2020}. Typical deep learning models, including both convolutional neural networks and GNNs, are designed so that they can characterize local features, and thus not extremely good at capturing non-local characteristics. The key to the present improvement is that the model learns relative motion between neighbor pairs to relate it to two-body structural correlation, in addition to the self-motion of the particles.  In this way, the GNN model bypasses
direct learning of non-local motion itself and acquires the ability to autonomously ``interpret'' how particle motion is affected by different dynamical effects non-local strain fluctuations and local particle rearrangements.  As a consequence, BOTAN reconciles the differences between fast $\beta$ and slow $\alpha$ relaxation dynamics and achieves state-of-the-art ability in the task of predicting the glassy dynamics from the static structure.

In the present study, particle propensity for motion defined on self displacement is chosen as the dynamical quantity of interest, in line with many previous studies on structure--dynamics correspondence in glasses~\cite{Harrowell2008,Hua2018,Tanaka2019,Bapst2020,Filion2021}. 
It was shown that the GNN model proposed by Bapst {\it et al.} (NT-GNN) can be reduced to a linear-regression (LR) model once properly-defined local structural parameters are averaged over neighbor shells, wherein predictive abilities are essentially the same between the LR model and NT-GNN~\cite{Filion2021}. Our results have, somewhat contrarily, shown that 
such local structural parameters are not the only descriptor for self-displacement, because they are under the influence of strain fluctuations. Recently, new neural network models that successfully improve the predictive ability on glassy dynamics by introducing different ideas have emerged — A recently proposed machine learning framework, GlassMLP, has raised the predictive ability by introducing the physics-informed descriptors’ input and bottleneck layers~\cite{Jung2022}; a couple of recent studies also seem to have improved the predictive ability of GNNs by introducing rotational-equivariant structural descriptors~\cite{Pezzicoli2022} or by additionally introducing self-attention layers into GNNs~\cite{Jiang2022}. As the reasons for the improved prediction are likely to be different from each other, it may be possible to develop a more advanced framework by combining these features, which is left as an open task in future. Given that a much higher prediction ability is realized, such models may also become useful for finding the ``reaction coordinates'' of glassy dynamics, along which the fluctuations should be enhanced for efficient sampling of molecular trajectories. Extensions to the recently discussed machine-learning-aided sampling technique may be interesting~\cite{Noe2019,Gabrie2022}.

\begin{acknowledgments}
We thank T. Kawasaki for critical input regarding this research. We also thank Y. Asahi, J. J. Molina, and F. Landes for helpful discussions and comments.  H.S. and T.Sh were supported by JSPS KAKENHI Grant Number JP19H05662 and by ``Joint Usage/Research Center for Interdisciplinary Large-scale Information Infrastructures'' and ``High Performance Computing Infrastructure'' in Japan (Project ID: jh220052). M.H. and T.Su were supported by ``Advanced Research Infrastructure for Materials and Nanotechnology'' from MEXT, Japan. Training and testing are performed using NVIDIA A100 Tensor Core GPUs (40GB SXM) on Wisteria/BDEC-01 Aquarius subsystem at Information Technology Center, University of Tokyo. For generation of the dataset for training and testing, usage of the following computational resources are acknowledged: Oakforest-PACS at the Joint Center for Advanced High Performance Computing (JCAHPC), Oakbridge-CX at Information Technology Center, University of Tokyo, Supercomputer System B at Institute for Solid State Physics, University of Tokyo, and MASAMUNE-IMR at Institute for Materials Research, Tohoku University.
\end{acknowledgments}

\section*{Data Availability Statement}
The source code used for training and evaluation is openly available at GitHub repository \url{https://github.com/h3-Open-BDEC/pyg_botan}. 
This code is an implementation of BOTAN using PyTorch Geometric~\cite{Fey/Lenssen/2019}. The dataset of simulation trajectories for evaluations will also be made available via the hyperlink indicated in the same GitHub repository for three years after the publication of this paper.
\appendix

\section{Effect of the iteration number in message passing} \label{sec:iteration}
\begin{figure}
\includegraphics[width=0.8\linewidth]{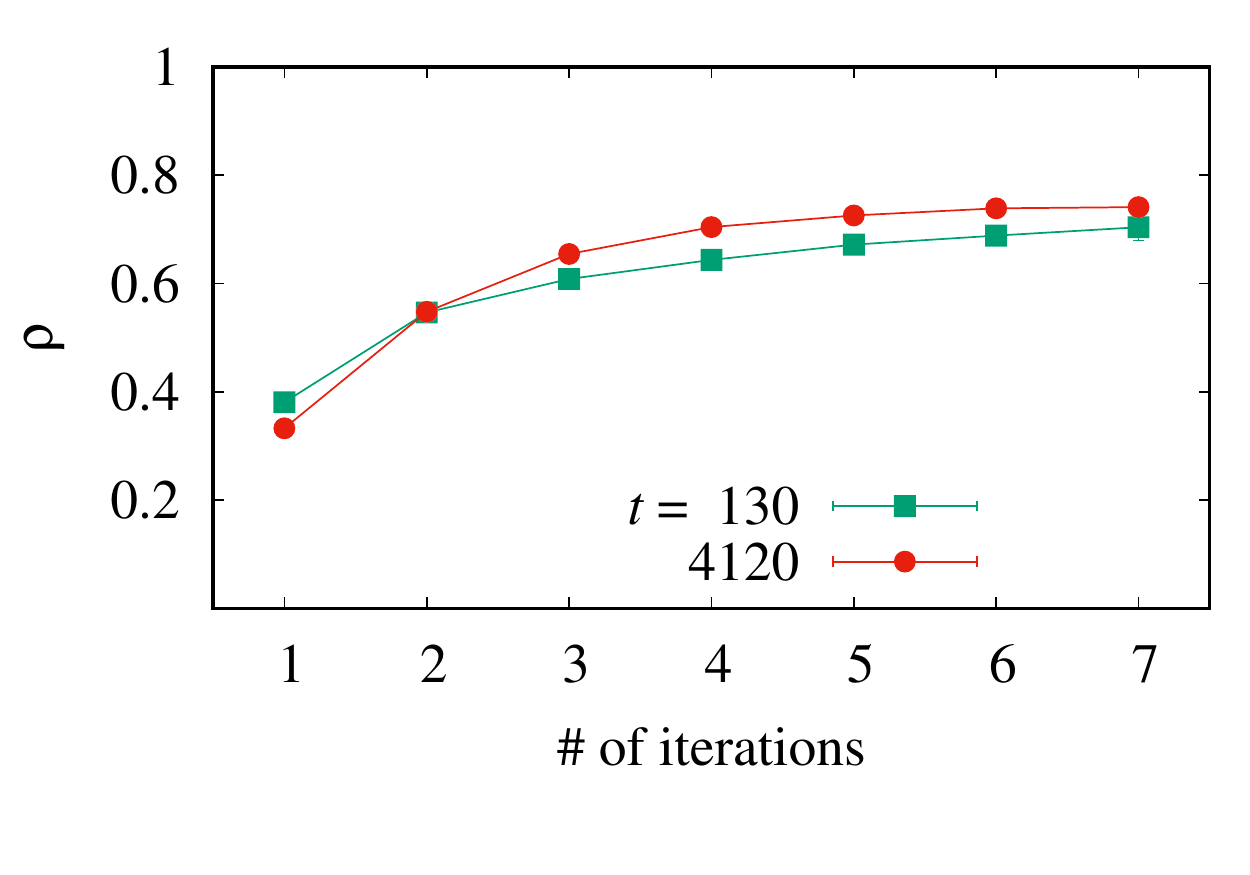}
\caption{\label{fig:iteration}
Pearson Correlation coefficients for the predictions of BOTAN at two time points, $t=130\ (0.03\tau_\alpha)$ and $4120\ (\tau_\alpha)$, at temperature $T=0.44$ are plotted as a function of the number of iterations of message passing.}
\end{figure}

In the model architecture of BOTAN, a pair of two-layer MLPs, respectively characterizing the features of nodes and edges, are mutually updated; edge updates are based on the features of the edge and its associated nodes, and node updates are based on the features of the node and on the sum of associated edge features, which the way of message passing between connected nodes in the present graph neural networks. Therefore, the number of repeat cycles determines from how far structural information can be incorporated into a node (or into an edge). In Fig.~\ref{fig:iteration}, the dependence of the Pearson correlation coefficient on this iteration number (denoted as $n$ in Sec.~\ref{sec:model}) is shown for $t=130$ and $4120$ at $T=0.44$, where the training is performed over 1000 epochs as well. The predictive ability increases as a function of the number of iterations $n$, but the increase is much slower when $n<4$.

\section{Optimization details related to training on edges} \label{sec:anticorr}
\begin{figure}
\includegraphics[width=0.8\linewidth]{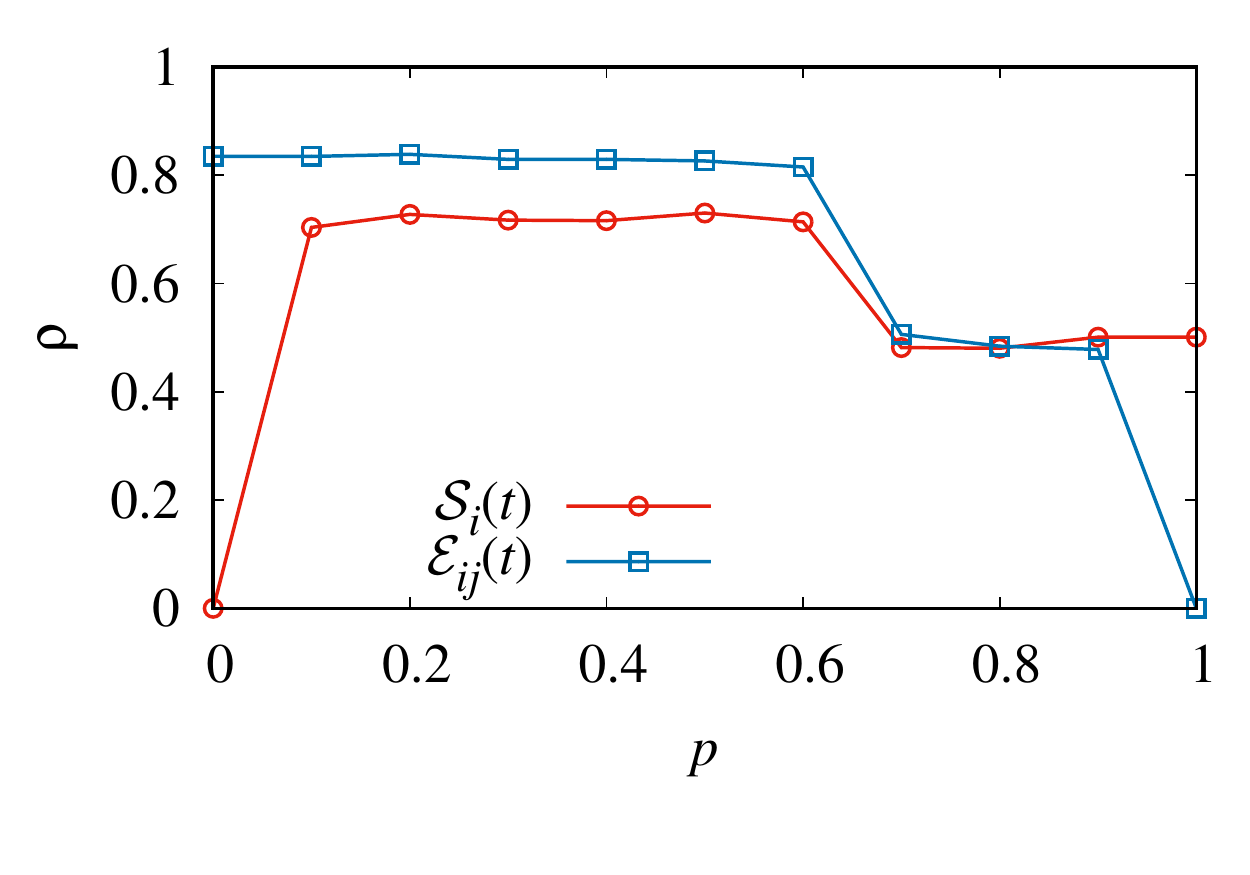}
\caption{\label{fig:pearson_bp}
Dependence of Pearson correlation coefficients for particle propensity $\mathcal{S}_i(t)$ and pair-distance changes $\mathcal{E}_{ij}(t)$ at $t=130$ and $T=0.44$, plotted as functions of $p$, a hyperparameter for the simultaneous learning. When $0<p<1$, these quantities are learned simultaneously. }
\end{figure}

In the simultaneous training on nodes and edges (via both $\mathcal{S}_i(t)$ and $\mathcal{E}_{ij}(t)$), the value of 0.4 is chosen for the hyperparameter $p$ that determines the training weight between nodes and edges.  This choice of the parameter is justified by an ablation experiment, in which BOTAN is trained for various values of the hyperparameter $p$ without using the pretrained model.  In Fig.~\ref{fig:pearson_bp}, we show Pearson correlation coefficients $\rho$ computed using predicted values for $\mathcal{S}_i(t)$ and $\mathcal{E}_{ij}(t)$, after being trained for 3000 epochs at $t=130$ and $T=0.44$.   The result suggests that $0.2 \le p\le 0.6$ may be a requirement that yields an optimal condition for the simultaneous training. 

\label{sec:ablation}

\bibliography{gnnglass,gnnglass_raw,revise}

\begin{thebibliography}{63}%
\makeatletter
\providecommand \@ifxundefined [1]{%
 \@ifx{#1\undefined}
}%
\providecommand \@ifnum [1]{%
 \ifnum #1\expandafter \@firstoftwo
 \else \expandafter \@secondoftwo
 \fi
}%
\providecommand \@ifx [1]{%
 \ifx #1\expandafter \@firstoftwo
 \else \expandafter \@secondoftwo
 \fi
}%
\providecommand \natexlab [1]{#1}%
\providecommand \enquote  [1]{``#1''}%
\providecommand \bibnamefont  [1]{#1}%
\providecommand \bibfnamefont [1]{#1}%
\providecommand \citenamefont [1]{#1}%
\providecommand \href@noop [0]{\@secondoftwo}%
\providecommand \href [0]{\begingroup \@sanitize@url \@href}%
\providecommand \@href[1]{\@@startlink{#1}\@@href}%
\providecommand \@@href[1]{\endgroup#1\@@endlink}%
\providecommand \@sanitize@url [0]{\catcode `\\12\catcode `\$12\catcode
  `\&12\catcode `\#12\catcode `\^12\catcode `\_12\catcode `\%12\relax}%
\providecommand \@@startlink[1]{}%
\providecommand \@@endlink[0]{}%
\providecommand \url  [0]{\begingroup\@sanitize@url \@url }%
\providecommand \@url [1]{\endgroup\@href {#1}{\urlprefix }}%
\providecommand \urlprefix  [0]{URL }%
\providecommand \Eprint [0]{\href }%
\providecommand \doibase [0]{http://dx.doi.org/}%
\providecommand \selectlanguage [0]{\@gobble}%
\providecommand \bibinfo  [0]{\@secondoftwo}%
\providecommand \bibfield  [0]{\@secondoftwo}%
\providecommand \translation [1]{[#1]}%
\providecommand \BibitemOpen [0]{}%
\providecommand \bibitemStop [0]{}%
\providecommand \bibitemNoStop [0]{.\EOS\space}%
\providecommand \EOS [0]{\spacefactor3000\relax}%
\providecommand \BibitemShut  [1]{\csname bibitem#1\endcsname}%
\let\auto@bib@innerbib\@empty
%</preamble>
\bibitem [{\citenamefont {Angell}(1995)}]{Angell1995}%
  \BibitemOpen
  \bibfield  {author} {\bibinfo {author} {\bibfnamefont {C.~A.}\ \bibnamefont
  {Angell}},\ }\bibfield  {title} {\enquote {\bibinfo {title} {Formation of
  glasses from liquids and biopolymers},}\ }\href {\doibase
  10.1126/science.267.5206.1924} {\bibfield  {journal} {\bibinfo  {journal}
  {Science}\ }\textbf {\bibinfo {volume} {267}},\ \bibinfo {pages} {1924--1935}
  (\bibinfo {year} {1995})}\BibitemShut {NoStop}%
\bibitem [{\citenamefont {Debenedetti}\ and\ \citenamefont
  {Stillinger}(2001)}]{Debenedetti2001}%
  \BibitemOpen
  \bibfield  {author} {\bibinfo {author} {\bibfnamefont {P.~G.}\ \bibnamefont
  {Debenedetti}}\ and\ \bibinfo {author} {\bibfnamefont {F.~H.}\ \bibnamefont
  {Stillinger}},\ }\bibfield  {title} {\enquote {\bibinfo {title} {Supercooled
  liquids and the glass transition},}\ }\href {www.nature.com} {\bibfield
  {journal} {\bibinfo  {journal} {Nature}\ }\textbf {\bibinfo {volume} {410}},\
  \bibinfo {pages} {259--267} (\bibinfo {year} {2001})}\BibitemShut {NoStop}%
\bibitem [{\citenamefont {Ediger}(2000)}]{Ediger2000}%
  \BibitemOpen
  \bibfield  {author} {\bibinfo {author} {\bibfnamefont {M.~D.}\ \bibnamefont
  {Ediger}},\ }\bibfield  {title} {\enquote {\bibinfo {title} {Spatially
  heterogeneous dynamics in supercooled liquids},}\ }\href {\doibase
  10.1146/annurev.physchem.51.1.99} {\bibfield  {journal} {\bibinfo  {journal}
  {Annu. Rev. Phys. Chem.}\ }\textbf {\bibinfo {volume} {51}},\ \bibinfo
  {pages} {99--128} (\bibinfo {year} {2000})}\BibitemShut {NoStop}%
\bibitem [{\citenamefont {Kob}\ \emph {et~al.}(1997)\citenamefont {Kob},
  \citenamefont {Donati}, \citenamefont {Plimpton}, \citenamefont {Poole},\
  and\ \citenamefont {Glotzer}}]{Kob1997}%
  \BibitemOpen
  \bibfield  {author} {\bibinfo {author} {\bibfnamefont {W.}~\bibnamefont
  {Kob}}, \bibinfo {author} {\bibfnamefont {C.}~\bibnamefont {Donati}},
  \bibinfo {author} {\bibfnamefont {S.~J.}\ \bibnamefont {Plimpton}}, \bibinfo
  {author} {\bibfnamefont {P.~H.}\ \bibnamefont {Poole}}, \ and\ \bibinfo
  {author} {\bibfnamefont {S.~C.}\ \bibnamefont {Glotzer}},\ }\bibfield
  {title} {\enquote {\bibinfo {title} {Dynamical heterogeneities in a
  supercooled lennard-jones liquid},}\ }\href@noop {} {\bibfield  {journal}
  {\bibinfo  {journal} {Physical Review Letters}\ }\textbf {\bibinfo {volume}
  {79}},\ \bibinfo {pages} {2827--2830} (\bibinfo {year} {1997})}\BibitemShut
  {NoStop}%
\bibitem [{\citenamefont {Yamamoto}\ and\ \citenamefont
  {Onuki}(1998)}]{Yamamoto1998b}%
  \BibitemOpen
  \bibfield  {author} {\bibinfo {author} {\bibfnamefont {R.}~\bibnamefont
  {Yamamoto}}\ and\ \bibinfo {author} {\bibfnamefont {A.}~\bibnamefont
  {Onuki}},\ }\bibfield  {title} {\enquote {\bibinfo {title} {Heterogeneous
  diffusion in highly supercooled liquids},}\ }\href@noop {} {\bibfield
  {journal} {\bibinfo  {journal} {Physical Review Letters}\ }\textbf {\bibinfo
  {volume} {81}},\ \bibinfo {pages} {4915--4918} (\bibinfo {year}
  {1998})}\BibitemShut {NoStop}%
\bibitem [{\citenamefont {Berthier}\ and\ \citenamefont
  {Biroli}(2011)}]{Berthier2011}%
  \BibitemOpen
  \bibfield  {author} {\bibinfo {author} {\bibfnamefont {L.}~\bibnamefont
  {Berthier}}\ and\ \bibinfo {author} {\bibfnamefont {G.}~\bibnamefont
  {Biroli}},\ }\bibfield  {title} {\enquote {\bibinfo {title} {Theoretical
  perspective on the glass transition and amorphous materials},}\ }\href
  {\doibase 10.1103/RevModPhys.83.587} {\bibfield  {journal} {\bibinfo
  {journal} {Reviews of Modern Physics}\ }\textbf {\bibinfo {volume} {83}},\
  \bibinfo {pages} {587--645} (\bibinfo {year} {2011})}\BibitemShut {NoStop}%
\bibitem [{\citenamefont {Tanaka}\ \emph {et~al.}(2019)\citenamefont {Tanaka},
  \citenamefont {Tong}, \citenamefont {Shi},\ and\ \citenamefont
  {Russo}}]{Tanaka2019}%
  \BibitemOpen
  \bibfield  {author} {\bibinfo {author} {\bibfnamefont {H.}~\bibnamefont
  {Tanaka}}, \bibinfo {author} {\bibfnamefont {H.}~\bibnamefont {Tong}},
  \bibinfo {author} {\bibfnamefont {R.}~\bibnamefont {Shi}}, \ and\ \bibinfo
  {author} {\bibfnamefont {J.}~\bibnamefont {Russo}},\ }\bibfield  {title}
  {\enquote {\bibinfo {title} {Revealing key structural features hidden in
  liquids and glasses},}\ }\href {\doibase 10.1038/s42254-019-0053-3}
  {\bibfield  {journal} {\bibinfo  {journal} {Nature Reviews Physics}\ }\textbf
  {\bibinfo {volume} {1}},\ \bibinfo {pages} {333--348} (\bibinfo {year}
  {2019})}\BibitemShut {NoStop}%
\bibitem [{\citenamefont {Royall}\ and\ \citenamefont
  {Williams}(2015)}]{Royall2015}%
  \BibitemOpen
  \bibfield  {author} {\bibinfo {author} {\bibfnamefont {C.~P.}\ \bibnamefont
  {Royall}}\ and\ \bibinfo {author} {\bibfnamefont {S.~R.}\ \bibnamefont
  {Williams}},\ }\bibfield  {title} {\enquote {\bibinfo {title} {The role of
  local structure in dynamical arrest},}\ }\href {\doibase
  10.1016/j.physrep.2014.11.004} {\bibfield  {journal} {\bibinfo  {journal}
  {Physics Reports}\ }\textbf {\bibinfo {volume} {560}},\ \bibinfo {pages}
  {1--75} (\bibinfo {year} {2015})}\BibitemShut {NoStop}%
\bibitem [{\citenamefont {Tong}\ and\ \citenamefont {Tanaka}(2018)}]{Hua2018}%
  \BibitemOpen
  \bibfield  {author} {\bibinfo {author} {\bibfnamefont {H.}~\bibnamefont
  {Tong}}\ and\ \bibinfo {author} {\bibfnamefont {H.}~\bibnamefont {Tanaka}},\
  }\bibfield  {title} {\enquote {\bibinfo {title} {Revealing hidden structural
  order controlling both fast and slow glassy dynamics in supercooled
  liquids},}\ }\href {\doibase 10.1103/PhysRevX.8.011041} {\bibfield  {journal}
  {\bibinfo  {journal} {Physical Review X}\ }\textbf {\bibinfo {volume} {8}},\
  \bibinfo {pages} {011041} (\bibinfo {year} {2018})}\BibitemShut {NoStop}%
\bibitem [{\citenamefont {Widmer-Cooper}\ \emph {et~al.}(2008)\citenamefont
  {Widmer-Cooper}, \citenamefont {Perry}, \citenamefont {Harrowell},\ and\
  \citenamefont {Reichman}}]{Harrowell2008}%
  \BibitemOpen
  \bibfield  {author} {\bibinfo {author} {\bibfnamefont {A.}~\bibnamefont
  {Widmer-Cooper}}, \bibinfo {author} {\bibfnamefont {H.}~\bibnamefont
  {Perry}}, \bibinfo {author} {\bibfnamefont {P.}~\bibnamefont {Harrowell}}, \
  and\ \bibinfo {author} {\bibfnamefont {D.~R.}\ \bibnamefont {Reichman}},\
  }\bibfield  {title} {\enquote {\bibinfo {title} {Irreversible reorganization
  in a supercooled liquid originates from localized soft modes},}\ }\href
  {\doibase 10.1038/nphys1025} {\bibfield  {journal} {\bibinfo  {journal}
  {Nature Physics}\ }\textbf {\bibinfo {volume} {4}},\ \bibinfo {pages}
  {711--715} (\bibinfo {year} {2008})}\BibitemShut {NoStop}%
\bibitem [{\citenamefont {Tanguy}, \citenamefont {Mantisi},\ and\ \citenamefont
  {Tsamados}(2010)}]{Tanguy2010}%
  \BibitemOpen
  \bibfield  {author} {\bibinfo {author} {\bibfnamefont {A.}~\bibnamefont
  {Tanguy}}, \bibinfo {author} {\bibfnamefont {B.}~\bibnamefont {Mantisi}}, \
  and\ \bibinfo {author} {\bibfnamefont {M.}~\bibnamefont {Tsamados}},\
  }\bibfield  {title} {\enquote {\bibinfo {title} {Vibrational modes as a
  predictor for plasticity in a model glass},}\ }\href {\doibase
  10.1209/0295-5075/90/16004} {\bibfield  {journal} {\bibinfo  {journal} {EPL}\
  }\textbf {\bibinfo {volume} {90}},\ \bibinfo {pages} {16004} (\bibinfo {year}
  {2010})}\BibitemShut {NoStop}%
\bibitem [{\citenamefont {Chen}\ \emph {et~al.}(2011)\citenamefont {Chen},
  \citenamefont {Manning}, \citenamefont {Yunker}, \citenamefont {Ellenbroek},
  \citenamefont {Zhang}, \citenamefont {Liu},\ and\ \citenamefont
  {Yodh}}]{Yodh2011}%
  \BibitemOpen
  \bibfield  {author} {\bibinfo {author} {\bibfnamefont {K.}~\bibnamefont
  {Chen}}, \bibinfo {author} {\bibfnamefont {M.~L.}\ \bibnamefont {Manning}},
  \bibinfo {author} {\bibfnamefont {P.~J.}\ \bibnamefont {Yunker}}, \bibinfo
  {author} {\bibfnamefont {W.~G.}\ \bibnamefont {Ellenbroek}}, \bibinfo
  {author} {\bibfnamefont {Z.}~\bibnamefont {Zhang}}, \bibinfo {author}
  {\bibfnamefont {A.~J.}\ \bibnamefont {Liu}}, \ and\ \bibinfo {author}
  {\bibfnamefont {A.~G.}\ \bibnamefont {Yodh}},\ }\bibfield  {title} {\enquote
  {\bibinfo {title} {Measurement of correlations between low-frequency
  vibrational modes and particle rearrangements in quasi-two-dimensional
  colloidal glasses},}\ }\href {\doibase 10.1103/PhysRevLett.107.108301}
  {\bibfield  {journal} {\bibinfo  {journal} {Physical Review Letters}\
  }\textbf {\bibinfo {volume} {107}},\ \bibinfo {pages} {108301} (\bibinfo
  {year} {2011})}\BibitemShut {NoStop}%
\bibitem [{\citenamefont {Ghosh}\ \emph {et~al.}(2011)\citenamefont {Ghosh},
  \citenamefont {Chikkadi}, \citenamefont {Schall},\ and\ \citenamefont
  {Bonn}}]{Bonn2011}%
  \BibitemOpen
  \bibfield  {author} {\bibinfo {author} {\bibfnamefont {A.}~\bibnamefont
  {Ghosh}}, \bibinfo {author} {\bibfnamefont {V.}~\bibnamefont {Chikkadi}},
  \bibinfo {author} {\bibfnamefont {P.}~\bibnamefont {Schall}}, \ and\ \bibinfo
  {author} {\bibfnamefont {D.}~\bibnamefont {Bonn}},\ }\bibfield  {title}
  {\enquote {\bibinfo {title} {Connecting structural relaxation with the low
  frequency modes in a hard-sphere colloidal glass},}\ }\href {\doibase
  10.1103/PhysRevLett.107.188303} {\bibfield  {journal} {\bibinfo  {journal}
  {Physical Review Letters}\ }\textbf {\bibinfo {volume} {107}},\ \bibinfo
  {pages} {188303} (\bibinfo {year} {2011})}\BibitemShut {NoStop}%
\bibitem [{\citenamefont {Schoenholz}\ \emph {et~al.}(2014)\citenamefont
  {Schoenholz}, \citenamefont {Liu}, \citenamefont {Riggleman},\ and\
  \citenamefont {Rottler}}]{Schoenholtz2014}%
  \BibitemOpen
  \bibfield  {author} {\bibinfo {author} {\bibfnamefont {S.~S.}\ \bibnamefont
  {Schoenholz}}, \bibinfo {author} {\bibfnamefont {A.~J.}\ \bibnamefont {Liu}},
  \bibinfo {author} {\bibfnamefont {R.~A.}\ \bibnamefont {Riggleman}}, \ and\
  \bibinfo {author} {\bibfnamefont {J.}~\bibnamefont {Rottler}},\ }\bibfield
  {title} {\enquote {\bibinfo {title} {Understanding plastic deformation in
  thermal glasses from single-soft-spot dynamics},}\ }\href {\doibase
  10.1103/PhysRevX.4.031014} {\bibfield  {journal} {\bibinfo  {journal}
  {Physical Review X}\ }\textbf {\bibinfo {volume} {4}},\ \bibinfo {pages}
  {031014} (\bibinfo {year} {2014})}\BibitemShut {NoStop}%
\bibitem [{\citenamefont {Ronhovde}\ \emph {et~al.}(2011)\citenamefont
  {Ronhovde}, \citenamefont {Chakrabarty}, \citenamefont {Hu}, \citenamefont
  {Sahu}, \citenamefont {Sahu}, \citenamefont {Kelton}, \citenamefont {Mauro},\
  and\ \citenamefont {Nussinov}}]{Nussinov2011}%
  \BibitemOpen
  \bibfield  {author} {\bibinfo {author} {\bibfnamefont {P.}~\bibnamefont
  {Ronhovde}}, \bibinfo {author} {\bibfnamefont {S.}~\bibnamefont
  {Chakrabarty}}, \bibinfo {author} {\bibfnamefont {D.}~\bibnamefont {Hu}},
  \bibinfo {author} {\bibfnamefont {M.}~\bibnamefont {Sahu}}, \bibinfo {author}
  {\bibfnamefont {K.~K.}\ \bibnamefont {Sahu}}, \bibinfo {author}
  {\bibfnamefont {K.~F.}\ \bibnamefont {Kelton}}, \bibinfo {author}
  {\bibfnamefont {N.~A.}\ \bibnamefont {Mauro}}, \ and\ \bibinfo {author}
  {\bibfnamefont {Z.}~\bibnamefont {Nussinov}},\ }\bibfield  {title} {\enquote
  {\bibinfo {title} {Detecting hidden spatial and spatio-temporal structures in
  glasses and complex physical systems by multiresolution network
  clustering},}\ }\href {\doibase 10.1140/epje/i2011-11105-9} {\bibfield
  {journal} {\bibinfo  {journal} {European Physical Journal E}\ }\textbf
  {\bibinfo {volume} {34}},\ \bibinfo {pages} {105} (\bibinfo {year}
  {2011})}\BibitemShut {NoStop}%
\bibitem [{\citenamefont {Cubuk}\ \emph {et~al.}(2015)\citenamefont {Cubuk},
  \citenamefont {Schoenholz}, \citenamefont {Rieser}, \citenamefont {Malone},
  \citenamefont {Rottler}, \citenamefont {Durian}, \citenamefont {Kaxiras},\
  and\ \citenamefont {Liu}}]{Cubuk2015}%
  \BibitemOpen
  \bibfield  {author} {\bibinfo {author} {\bibfnamefont {E.~D.}\ \bibnamefont
  {Cubuk}}, \bibinfo {author} {\bibfnamefont {S.~S.}\ \bibnamefont
  {Schoenholz}}, \bibinfo {author} {\bibfnamefont {J.~M.}\ \bibnamefont
  {Rieser}}, \bibinfo {author} {\bibfnamefont {B.~D.}\ \bibnamefont {Malone}},
  \bibinfo {author} {\bibfnamefont {J.}~\bibnamefont {Rottler}}, \bibinfo
  {author} {\bibfnamefont {D.~J.}\ \bibnamefont {Durian}}, \bibinfo {author}
  {\bibfnamefont {E.}~\bibnamefont {Kaxiras}}, \ and\ \bibinfo {author}
  {\bibfnamefont {A.~J.}\ \bibnamefont {Liu}},\ }\bibfield  {title} {\enquote
  {\bibinfo {title} {Identifying structural flow defects in disordered solids
  using machine-learning methods},}\ }\href {\doibase
  10.1103/PhysRevLett.114.108001} {\bibfield  {journal} {\bibinfo  {journal}
  {Physical Review Letters}\ }\textbf {\bibinfo {volume} {114}},\ \bibinfo
  {pages} {108001} (\bibinfo {year} {2015})}\BibitemShut {NoStop}%
\bibitem [{\citenamefont {Schoenholz}\ \emph {et~al.}(2016)\citenamefont
  {Schoenholz}, \citenamefont {Cubuk}, \citenamefont {Sussman}, \citenamefont
  {Kaxiras},\ and\ \citenamefont {Liu}}]{Schoenholtz2016}%
  \BibitemOpen
  \bibfield  {author} {\bibinfo {author} {\bibfnamefont {S.~S.}\ \bibnamefont
  {Schoenholz}}, \bibinfo {author} {\bibfnamefont {E.~D.}\ \bibnamefont
  {Cubuk}}, \bibinfo {author} {\bibfnamefont {D.~M.}\ \bibnamefont {Sussman}},
  \bibinfo {author} {\bibfnamefont {E.}~\bibnamefont {Kaxiras}}, \ and\
  \bibinfo {author} {\bibfnamefont {A.~J.}\ \bibnamefont {Liu}},\ }\bibfield
  {title} {\enquote {\bibinfo {title} {A structural approach to relaxation in
  glassy liquids},}\ }\href {\doibase 10.1038/nphys3644} {\bibfield  {journal}
  {\bibinfo  {journal} {Nature Physics}\ }\textbf {\bibinfo {volume} {12}},\
  \bibinfo {pages} {469--471} (\bibinfo {year} {2016})}\BibitemShut {NoStop}%
\bibitem [{\citenamefont {Schoenholz}\ \emph {et~al.}(2017)\citenamefont
  {Schoenholz}, \citenamefont {Cubuk}, \citenamefont {Kaxiras},\ and\
  \citenamefont {Liu}}]{Schoenholtz2017}%
  \BibitemOpen
  \bibfield  {author} {\bibinfo {author} {\bibfnamefont {S.~S.}\ \bibnamefont
  {Schoenholz}}, \bibinfo {author} {\bibfnamefont {E.~D.}\ \bibnamefont
  {Cubuk}}, \bibinfo {author} {\bibfnamefont {E.}~\bibnamefont {Kaxiras}}, \
  and\ \bibinfo {author} {\bibfnamefont {A.~J.}\ \bibnamefont {Liu}},\
  }\bibfield  {title} {\enquote {\bibinfo {title} {Relationship between local
  structure and relaxation in out-of-equilibrium glassy systems},}\ }\href
  {\doibase 10.1073/pnas.1610204114} {\bibfield  {journal} {\bibinfo  {journal}
  {Proceedings of the National Academy of Sciences of the United States of
  America}\ }\textbf {\bibinfo {volume} {114}},\ \bibinfo {pages} {263--267}
  (\bibinfo {year} {2017})}\BibitemShut {NoStop}%
\bibitem [{\citenamefont {Boattini}\ \emph {et~al.}(2020)\citenamefont
  {Boattini}, \citenamefont {Marín-Aguilar}, \citenamefont {Mitra},
  \citenamefont {Foffi}, \citenamefont {Smallenburg},\ and\ \citenamefont
  {Filion}}]{Filion2020}%
  \BibitemOpen
  \bibfield  {author} {\bibinfo {author} {\bibfnamefont {E.}~\bibnamefont
  {Boattini}}, \bibinfo {author} {\bibfnamefont {S.}~\bibnamefont
  {Marín-Aguilar}}, \bibinfo {author} {\bibfnamefont {S.}~\bibnamefont
  {Mitra}}, \bibinfo {author} {\bibfnamefont {G.}~\bibnamefont {Foffi}},
  \bibinfo {author} {\bibfnamefont {F.}~\bibnamefont {Smallenburg}}, \ and\
  \bibinfo {author} {\bibfnamefont {L.}~\bibnamefont {Filion}},\ }\bibfield
  {title} {\enquote {\bibinfo {title} {Autonomously revealing hidden local
  structures in supercooled liquids},}\ }\href {\doibase
  10.1038/s41467-020-19286-8} {\bibfield  {journal} {\bibinfo  {journal}
  {Nature Communications}\ }\textbf {\bibinfo {volume} {11}},\ \bibinfo {pages}
  {5479} (\bibinfo {year} {2020})}\BibitemShut {NoStop}%
\bibitem [{\citenamefont {Bapst}\ \emph {et~al.}(2020)\citenamefont {Bapst},
  \citenamefont {Keck}, \citenamefont {Grabska-Barwińska}, \citenamefont
  {Donner}, \citenamefont {Cubuk}, \citenamefont {Schoenholz}, \citenamefont
  {Obika}, \citenamefont {Nelson}, \citenamefont {Back}, \citenamefont
  {Hassabis},\ and\ \citenamefont {Kohli}}]{Bapst2020}%
  \BibitemOpen
  \bibfield  {author} {\bibinfo {author} {\bibfnamefont {V.}~\bibnamefont
  {Bapst}}, \bibinfo {author} {\bibfnamefont {T.}~\bibnamefont {Keck}},
  \bibinfo {author} {\bibfnamefont {A.}~\bibnamefont {Grabska-Barwińska}},
  \bibinfo {author} {\bibfnamefont {C.}~\bibnamefont {Donner}}, \bibinfo
  {author} {\bibfnamefont {E.~D.}\ \bibnamefont {Cubuk}}, \bibinfo {author}
  {\bibfnamefont {S.~S.}\ \bibnamefont {Schoenholz}}, \bibinfo {author}
  {\bibfnamefont {A.}~\bibnamefont {Obika}}, \bibinfo {author} {\bibfnamefont
  {A.~W.}\ \bibnamefont {Nelson}}, \bibinfo {author} {\bibfnamefont
  {T.}~\bibnamefont {Back}}, \bibinfo {author} {\bibfnamefont {D.}~\bibnamefont
  {Hassabis}}, \ and\ \bibinfo {author} {\bibfnamefont {P.}~\bibnamefont
  {Kohli}},\ }\bibfield  {title} {\enquote {\bibinfo {title} {Unveiling the
  predictive power of static structure in glassy systems},}\ }\href {\doibase
  10.1038/s41567-020-0842-8} {\bibfield  {journal} {\bibinfo  {journal} {Nature
  Physics}\ }\textbf {\bibinfo {volume} {16}},\ \bibinfo {pages} {448--454}
  (\bibinfo {year} {2020})}\BibitemShut {NoStop}%
\bibitem [{\citenamefont {Boattini}, \citenamefont {Smallenburg},\ and\
  \citenamefont {Filion}(2021)}]{Filion2021}%
  \BibitemOpen
  \bibfield  {author} {\bibinfo {author} {\bibfnamefont {E.}~\bibnamefont
  {Boattini}}, \bibinfo {author} {\bibfnamefont {F.}~\bibnamefont
  {Smallenburg}}, \ and\ \bibinfo {author} {\bibfnamefont {L.}~\bibnamefont
  {Filion}},\ }\bibfield  {title} {\enquote {\bibinfo {title} {Averaging local
  structure to predict the dynamic propensity in supercooled liquids},}\ }\href
  {\doibase 10.1103/PhysRevLett.127.088007} {\bibfield  {journal} {\bibinfo
  {journal} {Physical Review Letters}\ }\textbf {\bibinfo {volume} {127}},\
  \bibinfo {pages} {088007} (\bibinfo {year} {2021})}\BibitemShut {NoStop}%
\bibitem [{\citenamefont {Alkemade}\ \emph {et~al.}(2022)\citenamefont
  {Alkemade}, \citenamefont {Boattini}, \citenamefont {Filion},\ and\
  \citenamefont {Smallenburg}}]{Filion2022}%
  \BibitemOpen
  \bibfield  {author} {\bibinfo {author} {\bibfnamefont {R.~M.}\ \bibnamefont
  {Alkemade}}, \bibinfo {author} {\bibfnamefont {E.}~\bibnamefont {Boattini}},
  \bibinfo {author} {\bibfnamefont {L.}~\bibnamefont {Filion}}, \ and\ \bibinfo
  {author} {\bibfnamefont {F.}~\bibnamefont {Smallenburg}},\ }\bibfield
  {title} {\enquote {\bibinfo {title} {Comparing machine learning techniques
  for predicting glassy dynamics},}\ }\href {\doibase 10.1063/5.0088581}
  {\bibfield  {journal} {\bibinfo  {journal} {Journal of Chemical Physics}\
  }\textbf {\bibinfo {volume} {156}},\ \bibinfo {pages} {204503} (\bibinfo
  {year} {2022})}\BibitemShut {NoStop}%
\bibitem [{\citenamefont {Coslovich}, \citenamefont {Jack},\ and\ \citenamefont
  {Paret}(2022)}]{Coslovich2022}%
  \BibitemOpen
  \bibfield  {author} {\bibinfo {author} {\bibfnamefont {D.}~\bibnamefont
  {Coslovich}}, \bibinfo {author} {\bibfnamefont {R.~L.}\ \bibnamefont {Jack}},
  \ and\ \bibinfo {author} {\bibfnamefont {J.}~\bibnamefont {Paret}},\
  }\bibfield  {title} {\enquote {\bibinfo {title} {Dimensionality reduction of
  local structure in glassy binary mixtures},}\ }\href {\doibase
  10.1063/5.0128265} {\bibfield  {journal} {\bibinfo  {journal} {J. Chem.
  Phys,}\ }\textbf {\bibinfo {volume} {157}},\ \bibinfo {pages} {204503}
  (\bibinfo {year} {2022})}\BibitemShut {NoStop}%
\bibitem [{\citenamefont {Scarselli}\ \emph {et~al.}(2009)\citenamefont
  {Scarselli}, \citenamefont {Gori}, \citenamefont {Tsoi}, \citenamefont
  {Hagenbuchner},\ and\ \citenamefont {Monfardini}}]{Scarselli2009}%
  \BibitemOpen
  \bibfield  {author} {\bibinfo {author} {\bibfnamefont {F.}~\bibnamefont
  {Scarselli}}, \bibinfo {author} {\bibfnamefont {M.}~\bibnamefont {Gori}},
  \bibinfo {author} {\bibfnamefont {A.~C.}\ \bibnamefont {Tsoi}}, \bibinfo
  {author} {\bibfnamefont {M.}~\bibnamefont {Hagenbuchner}}, \ and\ \bibinfo
  {author} {\bibfnamefont {G.}~\bibnamefont {Monfardini}},\ }\bibfield  {title}
  {\enquote {\bibinfo {title} {The graph neural network model},}\ }\href
  {\doibase 10.1109/TNN.2008.2005605} {\bibfield  {journal} {\bibinfo
  {journal} {IEEE Transactions on Neural Networks}\ }\textbf {\bibinfo {volume}
  {20}},\ \bibinfo {pages} {61--80} (\bibinfo {year} {2009})}\BibitemShut
  {NoStop}%
\bibitem [{\citenamefont {Xie}\ and\ \citenamefont
  {Grossman}(2018)}]{CGCNN_2018}%
  \BibitemOpen
  \bibfield  {author} {\bibinfo {author} {\bibfnamefont {T.}~\bibnamefont
  {Xie}}\ and\ \bibinfo {author} {\bibfnamefont {J.~C.}\ \bibnamefont
  {Grossman}},\ }\bibfield  {title} {\enquote {\bibinfo {title} {Crystal graph
  convolutional neural networks for an accurate and interpretable prediction of
  material properties},}\ }\href {\doibase 10.1103/PhysRevLett.120.145301}
  {\bibfield  {journal} {\bibinfo  {journal} {Physical Review Letters}\
  }\textbf {\bibinfo {volume} {120}},\ \bibinfo {pages} {145301} (\bibinfo
  {year} {2018})}\BibitemShut {NoStop}%
\bibitem [{\citenamefont {Chanussot}\ \emph {et~al.}(2021)\citenamefont
  {Chanussot}, \citenamefont {Das}, \citenamefont {Goyal}, \citenamefont
  {Lavril}, \citenamefont {Shuaibi}, \citenamefont {Riviere}, \citenamefont
  {Tran}, \citenamefont {Heras-Domingo}, \citenamefont {Ho}, \citenamefont
  {Hu}, \citenamefont {Palizhati}, \citenamefont {Sriram}, \citenamefont
  {Wood}, \citenamefont {Yoon}, \citenamefont {Parikh}, \citenamefont
  {Zitnick},\ and\ \citenamefont {Ulissi}}]{OpenCatalyst2021}%
  \BibitemOpen
  \bibfield  {author} {\bibinfo {author} {\bibfnamefont {L.}~\bibnamefont
  {Chanussot}}, \bibinfo {author} {\bibfnamefont {A.}~\bibnamefont {Das}},
  \bibinfo {author} {\bibfnamefont {S.}~\bibnamefont {Goyal}}, \bibinfo
  {author} {\bibfnamefont {T.}~\bibnamefont {Lavril}}, \bibinfo {author}
  {\bibfnamefont {M.}~\bibnamefont {Shuaibi}}, \bibinfo {author} {\bibfnamefont
  {M.}~\bibnamefont {Riviere}}, \bibinfo {author} {\bibfnamefont
  {K.}~\bibnamefont {Tran}}, \bibinfo {author} {\bibfnamefont {J.}~\bibnamefont
  {Heras-Domingo}}, \bibinfo {author} {\bibfnamefont {C.}~\bibnamefont {Ho}},
  \bibinfo {author} {\bibfnamefont {W.}~\bibnamefont {Hu}}, \bibinfo {author}
  {\bibfnamefont {A.}~\bibnamefont {Palizhati}}, \bibinfo {author}
  {\bibfnamefont {A.}~\bibnamefont {Sriram}}, \bibinfo {author} {\bibfnamefont
  {B.}~\bibnamefont {Wood}}, \bibinfo {author} {\bibfnamefont {J.}~\bibnamefont
  {Yoon}}, \bibinfo {author} {\bibfnamefont {D.}~\bibnamefont {Parikh}},
  \bibinfo {author} {\bibfnamefont {C.~L.}\ \bibnamefont {Zitnick}}, \ and\
  \bibinfo {author} {\bibfnamefont {Z.}~\bibnamefont {Ulissi}},\ }\bibfield
  {title} {\enquote {\bibinfo {title} {Open catalyst 2020 (oc20) dataset and
  community challenges},}\ }\href {\doibase 10.1021/acscatal.0c04525}
  {\bibfield  {journal} {\bibinfo  {journal} {ACS Catalysis}\ }\textbf
  {\bibinfo {volume} {11}},\ \bibinfo {pages} {6059--6072} (\bibinfo {year}
  {2021})}\BibitemShut {NoStop}%
\bibitem [{\citenamefont {Fung}\ \emph {et~al.}(2021)\citenamefont {Fung},
  \citenamefont {Zhang}, \citenamefont {Juarez},\ and\ \citenamefont
  {Sumpter}}]{Fung2021}%
  \BibitemOpen
  \bibfield  {author} {\bibinfo {author} {\bibfnamefont {V.}~\bibnamefont
  {Fung}}, \bibinfo {author} {\bibfnamefont {J.}~\bibnamefont {Zhang}},
  \bibinfo {author} {\bibfnamefont {E.}~\bibnamefont {Juarez}}, \ and\ \bibinfo
  {author} {\bibfnamefont {B.~G.}\ \bibnamefont {Sumpter}},\ }\bibfield
  {title} {\enquote {\bibinfo {title} {Benchmarking graph neural networks for
  materials chemistry},}\ }\href {\doibase 10.1038/s41524-021-00554-0}
  {\bibfield  {journal} {\bibinfo  {journal} {npj Computational Materials}\
  }\textbf {\bibinfo {volume} {7}},\ \bibinfo {pages} {84} (\bibinfo {year}
  {2021})}\BibitemShut {NoStop}%
\bibitem [{\citenamefont {Takamoto}\ \emph {et~al.}(2022)\citenamefont
  {Takamoto}, \citenamefont {Shinagawa}, \citenamefont {Motoki}, \citenamefont
  {Nakago}, \citenamefont {Li}, \citenamefont {Kurata}, \citenamefont
  {Watanabe}, \citenamefont {Yayama}, \citenamefont {Iriguchi}, \citenamefont
  {Asano}, \citenamefont {Onodera}, \citenamefont {Ishii}, \citenamefont
  {Kudo}, \citenamefont {Ono}, \citenamefont {Sawada}, \citenamefont
  {Ishitani}, \citenamefont {Ong}, \citenamefont {Yamaguchi}, \citenamefont
  {Kataoka}, \citenamefont {Hayashi}, \citenamefont {Charoenphakdee},\ and\
  \citenamefont {Ibuka}}]{PFN2022}%
  \BibitemOpen
  \bibfield  {author} {\bibinfo {author} {\bibfnamefont {S.}~\bibnamefont
  {Takamoto}}, \bibinfo {author} {\bibfnamefont {C.}~\bibnamefont {Shinagawa}},
  \bibinfo {author} {\bibfnamefont {D.}~\bibnamefont {Motoki}}, \bibinfo
  {author} {\bibfnamefont {K.}~\bibnamefont {Nakago}}, \bibinfo {author}
  {\bibfnamefont {W.}~\bibnamefont {Li}}, \bibinfo {author} {\bibfnamefont
  {I.}~\bibnamefont {Kurata}}, \bibinfo {author} {\bibfnamefont
  {T.}~\bibnamefont {Watanabe}}, \bibinfo {author} {\bibfnamefont
  {Y.}~\bibnamefont {Yayama}}, \bibinfo {author} {\bibfnamefont
  {H.}~\bibnamefont {Iriguchi}}, \bibinfo {author} {\bibfnamefont
  {Y.}~\bibnamefont {Asano}}, \bibinfo {author} {\bibfnamefont
  {T.}~\bibnamefont {Onodera}}, \bibinfo {author} {\bibfnamefont
  {T.}~\bibnamefont {Ishii}}, \bibinfo {author} {\bibfnamefont
  {T.}~\bibnamefont {Kudo}}, \bibinfo {author} {\bibfnamefont {H.}~\bibnamefont
  {Ono}}, \bibinfo {author} {\bibfnamefont {R.}~\bibnamefont {Sawada}},
  \bibinfo {author} {\bibfnamefont {R.}~\bibnamefont {Ishitani}}, \bibinfo
  {author} {\bibfnamefont {M.}~\bibnamefont {Ong}}, \bibinfo {author}
  {\bibfnamefont {T.}~\bibnamefont {Yamaguchi}}, \bibinfo {author}
  {\bibfnamefont {T.}~\bibnamefont {Kataoka}}, \bibinfo {author} {\bibfnamefont
  {A.}~\bibnamefont {Hayashi}}, \bibinfo {author} {\bibfnamefont
  {N.}~\bibnamefont {Charoenphakdee}}, \ and\ \bibinfo {author} {\bibfnamefont
  {T.}~\bibnamefont {Ibuka}},\ }\bibfield  {title} {\enquote {\bibinfo {title}
  {Towards universal neural network potential for material discovery applicable
  to arbitrary combination of 45 elements},}\ }\href {\doibase
  10.1038/s41467-022-30687-9} {\bibfield  {journal} {\bibinfo  {journal}
  {Nature Communications}\ }\textbf {\bibinfo {volume} {13}},\ \bibinfo {pages}
  {2991} (\bibinfo {year} {2022})}\BibitemShut {NoStop}%
\bibitem [{\citenamefont {Wang}\ and\ \citenamefont {Zhang}(2021)}]{Wang2021}%
  \BibitemOpen
  \bibfield  {author} {\bibinfo {author} {\bibfnamefont {Q.}~\bibnamefont
  {Wang}}\ and\ \bibinfo {author} {\bibfnamefont {L.}~\bibnamefont {Zhang}},\
  }\bibfield  {title} {\enquote {\bibinfo {title} {Inverse design of glass
  structure with deep graph neural networks},}\ }\href {\doibase
  10.1038/s41467-021-25490-x} {\bibfield  {journal} {\bibinfo  {journal}
  {Nature Communications}\ }\textbf {\bibinfo {volume} {12}},\ \bibinfo {pages}
  {5359} (\bibinfo {year} {2021})}\BibitemShut {NoStop}%
\bibitem [{\citenamefont {Buchenau}\ and\ \citenamefont
  {Zorn}(1992)}]{Buchenau1992}%
  \BibitemOpen
  \bibfield  {author} {\bibinfo {author} {\bibfnamefont {U.}~\bibnamefont
  {Buchenau}}\ and\ \bibinfo {author} {\bibfnamefont {R.}~\bibnamefont
  {Zorn}},\ }\bibfield  {title} {\enquote {\bibinfo {title} {A relation between
  fast and slow motions in glassy and liquid selenium},}\ }\href@noop {}
  {\bibfield  {journal} {\bibinfo  {journal} {EPL}\ }\textbf {\bibinfo {volume}
  {18}},\ \bibinfo {pages} {523--528} (\bibinfo {year} {1992})}\BibitemShut
  {NoStop}%
\bibitem [{\citenamefont {Widmer-Cooper}\ and\ \citenamefont
  {Harrowell}(2006)}]{Widmer-Cooper2006}%
  \BibitemOpen
  \bibfield  {author} {\bibinfo {author} {\bibfnamefont {A.}~\bibnamefont
  {Widmer-Cooper}}\ and\ \bibinfo {author} {\bibfnamefont {P.}~\bibnamefont
  {Harrowell}},\ }\bibfield  {title} {\enquote {\bibinfo {title} {Predicting
  the long-time dynamic heterogeneity in a supercooled liquid on the basis of
  short-time heterogeneities},}\ }\href {\doibase
  10.1103/PhysRevLett.96.185701} {\bibfield  {journal} {\bibinfo  {journal}
  {Physical Review Letters}\ }\textbf {\bibinfo {volume} {96}},\ \bibinfo
  {pages} {185701} (\bibinfo {year} {2006})}\BibitemShut {NoStop}%
\bibitem [{\citenamefont {Dyre}(2006)}]{Dyre2006}%
  \BibitemOpen
  \bibfield  {author} {\bibinfo {author} {\bibfnamefont {J.~C.}\ \bibnamefont
  {Dyre}},\ }\bibfield  {title} {\enquote {\bibinfo {title} {Colloquium: The
  glass transition and elastic models of glass-forming liquids},}\ }\href
  {\doibase 10.1103/RevModPhys.78.953} {\bibfield  {journal} {\bibinfo
  {journal} {Reviews of Modern Physics}\ }\textbf {\bibinfo {volume} {78}},\
  \bibinfo {pages} {953--972} (\bibinfo {year} {2006})}\BibitemShut {NoStop}%
\bibitem [{\citenamefont {Larini}\ \emph {et~al.}(2008)\citenamefont {Larini},
  \citenamefont {Ottochian}, \citenamefont {Michele},\ and\ \citenamefont
  {Leporini}}]{Leporini2016}%
  \BibitemOpen
  \bibfield  {author} {\bibinfo {author} {\bibfnamefont {L.}~\bibnamefont
  {Larini}}, \bibinfo {author} {\bibfnamefont {A.}~\bibnamefont {Ottochian}},
  \bibinfo {author} {\bibfnamefont {C.~D.}\ \bibnamefont {Michele}}, \ and\
  \bibinfo {author} {\bibfnamefont {D.}~\bibnamefont {Leporini}},\ }\bibfield
  {title} {\enquote {\bibinfo {title} {Universal scaling between structural
  relaxation and vibrational dynamics inglass-forming liquids and polymers},}\
  }\href {\doibase 10.1038/nphys788} {\bibfield  {journal} {\bibinfo  {journal}
  {Nature Physics}\ }\textbf {\bibinfo {volume} {4}},\ \bibinfo {pages}
  {42--45} (\bibinfo {year} {2008})}\BibitemShut {NoStop}%
\bibitem [{\citenamefont {Hansen}\ \emph {et~al.}(2017)\citenamefont {Hansen},
  \citenamefont {Frick}, \citenamefont {Hecksher}, \citenamefont {Dyre},\ and\
  \citenamefont {Niss}}]{Dyre2017}%
  \BibitemOpen
  \bibfield  {author} {\bibinfo {author} {\bibfnamefont {H.~W.}\ \bibnamefont
  {Hansen}}, \bibinfo {author} {\bibfnamefont {B.}~\bibnamefont {Frick}},
  \bibinfo {author} {\bibfnamefont {T.}~\bibnamefont {Hecksher}}, \bibinfo
  {author} {\bibfnamefont {J.~C.}\ \bibnamefont {Dyre}}, \ and\ \bibinfo
  {author} {\bibfnamefont {K.}~\bibnamefont {Niss}},\ }\bibfield  {title}
  {\enquote {\bibinfo {title} {Connection between fragility, mean-squared
  displacement, and shear modulus in two van der waals bonded glass-forming
  liquids},}\ }\href {\doibase 10.1103/PhysRevB.95.104202} {\bibfield
  {journal} {\bibinfo  {journal} {Physical Review B}\ }\textbf {\bibinfo
  {volume} {95}},\ \bibinfo {pages} {104202} (\bibinfo {year}
  {2017})}\BibitemShut {NoStop}%
\bibitem [{\citenamefont {Widmer-Cooper}\ \emph {et~al.}(2009)\citenamefont
  {Widmer-Cooper}, \citenamefont {Perry}, \citenamefont {Harrowell},\ and\
  \citenamefont {Reichman}}]{Widmer2009}%
  \BibitemOpen
  \bibfield  {author} {\bibinfo {author} {\bibfnamefont {A.}~\bibnamefont
  {Widmer-Cooper}}, \bibinfo {author} {\bibfnamefont {H.}~\bibnamefont
  {Perry}}, \bibinfo {author} {\bibfnamefont {P.}~\bibnamefont {Harrowell}}, \
  and\ \bibinfo {author} {\bibfnamefont {D.~R.}\ \bibnamefont {Reichman}},\
  }\bibfield  {title} {\enquote {\bibinfo {title} {Localized soft modes and the
  supercooled liquid's irreversible passage through its configuration space},}\
  }\href {\doibase 10.1063/1.3265983} {\bibfield  {journal} {\bibinfo
  {journal} {Journal of Chemical Physics}\ }\textbf {\bibinfo {volume} {131}},\
  \bibinfo {pages} {194508} (\bibinfo {year} {2009})}\BibitemShut {NoStop}%
\bibitem [{\citenamefont {Tripodo}\ \emph {et~al.}(2022)\citenamefont
  {Tripodo}, \citenamefont {Cordella}, \citenamefont {Puosi}, \citenamefont
  {Malvaldi},\ and\ \citenamefont {Leporini}}]{Puosi2022}%
  \BibitemOpen
  \bibfield  {author} {\bibinfo {author} {\bibfnamefont {A.}~\bibnamefont
  {Tripodo}}, \bibinfo {author} {\bibfnamefont {G.}~\bibnamefont {Cordella}},
  \bibinfo {author} {\bibfnamefont {F.}~\bibnamefont {Puosi}}, \bibinfo
  {author} {\bibfnamefont {M.}~\bibnamefont {Malvaldi}}, \ and\ \bibinfo
  {author} {\bibfnamefont {D.}~\bibnamefont {Leporini}},\ }\bibfield  {title}
  {\enquote {\bibinfo {title} {Neural networks reveal the impact of the
  vibrational dynamics in the prediction of the long-time mobility of molecular
  glassformers},}\ }\href {\doibase 10.3390/ijms23169322} {\bibfield  {journal}
  {\bibinfo  {journal} {International Journal of Molecular Sciences}\ }\textbf
  {\bibinfo {volume} {23}},\ \bibinfo {pages} {9322} (\bibinfo {year}
  {2022})}\BibitemShut {NoStop}%
\bibitem [{\citenamefont {Battaglia}\ \emph {et~al.}(2016)\citenamefont
  {Battaglia}, \citenamefont {Pascanu}, \citenamefont {Lai}, \citenamefont
  {Rezende},\ and\ \citenamefont {Kavukcuoglu}}]{Battaglia2016}%
  \BibitemOpen
  \bibfield  {author} {\bibinfo {author} {\bibfnamefont {P.~W.}\ \bibnamefont
  {Battaglia}}, \bibinfo {author} {\bibfnamefont {R.}~\bibnamefont {Pascanu}},
  \bibinfo {author} {\bibfnamefont {M.}~\bibnamefont {Lai}}, \bibinfo {author}
  {\bibfnamefont {D.}~\bibnamefont {Rezende}}, \ and\ \bibinfo {author}
  {\bibfnamefont {K.}~\bibnamefont {Kavukcuoglu}},\ }\bibfield  {title}
  {\enquote {\bibinfo {title} {Interaction networks for learning about objects,
  relations and physics},}\ }\href {http://arxiv.org/abs/1612.00222} {\bibfield
   {journal} {\bibinfo  {journal} {Advances in Neural Information Processing
  Systems}\ }\textbf {\bibinfo {volume} {29}},\ \bibinfo {pages} {4502--4510}
  (\bibinfo {year} {2016})}\BibitemShut {NoStop}%
\bibitem [{\citenamefont {Battaglia}\ \emph {et~al.}(2018)\citenamefont
  {Battaglia}, \citenamefont {Hamrick}, \citenamefont {Bapst}, \citenamefont
  {Sanchez-Gonzalez}, \citenamefont {Zambaldi}, \citenamefont {Malinowski},
  \citenamefont {Tacchetti}, \citenamefont {Raposo}, \citenamefont {Santoro},
  \citenamefont {Faulkner}, \citenamefont {Gulcehre}, \citenamefont {Song},
  \citenamefont {Ballard}, \citenamefont {Gilmer}, \citenamefont {Dahl},
  \citenamefont {Vaswani}, \citenamefont {Allen}, \citenamefont {Nash},
  \citenamefont {Langston}, \citenamefont {Dyer}, \citenamefont {Heess},
  \citenamefont {Wierstra}, \citenamefont {Kohli}, \citenamefont {Botvinick},
  \citenamefont {Vinyals}, \citenamefont {Li},\ and\ \citenamefont
  {Pascanu}}]{Battaglia2018}%
  \BibitemOpen
  \bibfield  {author} {\bibinfo {author} {\bibfnamefont {P.~W.}\ \bibnamefont
  {Battaglia}}, \bibinfo {author} {\bibfnamefont {J.~B.}\ \bibnamefont
  {Hamrick}}, \bibinfo {author} {\bibfnamefont {V.}~\bibnamefont {Bapst}},
  \bibinfo {author} {\bibfnamefont {A.}~\bibnamefont {Sanchez-Gonzalez}},
  \bibinfo {author} {\bibfnamefont {V.}~\bibnamefont {Zambaldi}}, \bibinfo
  {author} {\bibfnamefont {M.}~\bibnamefont {Malinowski}}, \bibinfo {author}
  {\bibfnamefont {A.}~\bibnamefont {Tacchetti}}, \bibinfo {author}
  {\bibfnamefont {D.}~\bibnamefont {Raposo}}, \bibinfo {author} {\bibfnamefont
  {A.}~\bibnamefont {Santoro}}, \bibinfo {author} {\bibfnamefont
  {R.}~\bibnamefont {Faulkner}}, \bibinfo {author} {\bibfnamefont
  {C.}~\bibnamefont {Gulcehre}}, \bibinfo {author} {\bibfnamefont
  {F.}~\bibnamefont {Song}}, \bibinfo {author} {\bibfnamefont {A.}~\bibnamefont
  {Ballard}}, \bibinfo {author} {\bibfnamefont {J.}~\bibnamefont {Gilmer}},
  \bibinfo {author} {\bibfnamefont {G.}~\bibnamefont {Dahl}}, \bibinfo {author}
  {\bibfnamefont {A.}~\bibnamefont {Vaswani}}, \bibinfo {author} {\bibfnamefont
  {K.}~\bibnamefont {Allen}}, \bibinfo {author} {\bibfnamefont
  {C.}~\bibnamefont {Nash}}, \bibinfo {author} {\bibfnamefont {V.}~\bibnamefont
  {Langston}}, \bibinfo {author} {\bibfnamefont {C.}~\bibnamefont {Dyer}},
  \bibinfo {author} {\bibfnamefont {N.}~\bibnamefont {Heess}}, \bibinfo
  {author} {\bibfnamefont {D.}~\bibnamefont {Wierstra}}, \bibinfo {author}
  {\bibfnamefont {P.}~\bibnamefont {Kohli}}, \bibinfo {author} {\bibfnamefont
  {M.}~\bibnamefont {Botvinick}}, \bibinfo {author} {\bibfnamefont
  {O.}~\bibnamefont {Vinyals}}, \bibinfo {author} {\bibfnamefont
  {Y.}~\bibnamefont {Li}}, \ and\ \bibinfo {author} {\bibfnamefont
  {R.}~\bibnamefont {Pascanu}},\ }\href@noop {} {\enquote {\bibinfo {title}
  {Relational inductive biases, deep learning, and graph networks},}\ }
  (\bibinfo {year} {2018}),\ \Eprint {http://arxiv.org/abs/1806.01261}
  {arXiv:1806.01261 [cs.LG]} \BibitemShut {NoStop}%
\bibitem [{\citenamefont {DeZoort}\ \emph {et~al.}(2021)\citenamefont
  {DeZoort}, \citenamefont {Thais}, \citenamefont {Duarte}, \citenamefont
  {Razavimaleki}, \citenamefont {Atkinson}, \citenamefont {Ojalvo},
  \citenamefont {Neubauer},\ and\ \citenamefont {Elmer}}]{DeZoort2021}%
  \BibitemOpen
  \bibfield  {author} {\bibinfo {author} {\bibfnamefont {G.}~\bibnamefont
  {DeZoort}}, \bibinfo {author} {\bibfnamefont {S.}~\bibnamefont {Thais}},
  \bibinfo {author} {\bibfnamefont {J.}~\bibnamefont {Duarte}}, \bibinfo
  {author} {\bibfnamefont {V.}~\bibnamefont {Razavimaleki}}, \bibinfo {author}
  {\bibfnamefont {M.}~\bibnamefont {Atkinson}}, \bibinfo {author}
  {\bibfnamefont {I.}~\bibnamefont {Ojalvo}}, \bibinfo {author} {\bibfnamefont
  {M.}~\bibnamefont {Neubauer}}, \ and\ \bibinfo {author} {\bibfnamefont
  {P.}~\bibnamefont {Elmer}},\ }\bibfield  {title} {\enquote {\bibinfo {title}
  {Charged particle tracking via edge-classifying interaction networks},}\
  }\href {\doibase 10.1007/s41781-021-00073-z} {\bibfield  {journal} {\bibinfo
  {journal} {Computing and Software for Big Science}\ }\textbf {\bibinfo
  {volume} {5}},\ \bibinfo {pages} {26} (\bibinfo {year} {2021})}\BibitemShut
  {NoStop}%
\bibitem [{\citenamefont {Kob}\ and\ \citenamefont {Andersen}(1995)}]{Kob1995}%
  \BibitemOpen
  \bibfield  {author} {\bibinfo {author} {\bibfnamefont {W.}~\bibnamefont
  {Kob}}\ and\ \bibinfo {author} {\bibfnamefont {H.~C.}\ \bibnamefont
  {Andersen}},\ }\bibfield  {title} {\enquote {\bibinfo {title} {Testing
  mode-coupling theory for a supercooled binary lennard-jones mixture: The van
  hove correlation function},}\ }\href {\doibase 10.1103/PhysRevE.51.4626}
  {\bibfield  {journal} {\bibinfo  {journal} {Physical Review E}\ }\textbf
  {\bibinfo {volume} {51}},\ \bibinfo {pages} {4626--4641} (\bibinfo {year}
  {1995})}\BibitemShut {NoStop}%
\bibitem [{\citenamefont {Shimada}, \citenamefont {Mizuno},\ and\ \citenamefont
  {Ikeda}(2018)}]{Shimada2018}%
  \BibitemOpen
  \bibfield  {author} {\bibinfo {author} {\bibfnamefont {M.}~\bibnamefont
  {Shimada}}, \bibinfo {author} {\bibfnamefont {H.}~\bibnamefont {Mizuno}}, \
  and\ \bibinfo {author} {\bibfnamefont {A.}~\bibnamefont {Ikeda}},\ }\bibfield
   {title} {\enquote {\bibinfo {title} {Anomalous vibrational properties in the
  continuum limit of glasses},}\ }\href {\doibase 10.1103/PhysRevE.97.022609}
  {\bibfield  {journal} {\bibinfo  {journal} {Physical Review E}\ }\textbf
  {\bibinfo {volume} {97}},\ \bibinfo {pages} {022609} (\bibinfo {year}
  {2018})}\BibitemShut {NoStop}%
\bibitem [{\citenamefont {Toxvaerd}\ and\ \citenamefont
  {Dyre}(2011)}]{Dyre2011}%
  \BibitemOpen
  \bibfield  {author} {\bibinfo {author} {\bibfnamefont {S.}~\bibnamefont
  {Toxvaerd}}\ and\ \bibinfo {author} {\bibfnamefont {J.~C.}\ \bibnamefont
  {Dyre}},\ }\bibfield  {title} {\enquote {\bibinfo {title} {Communication:
  Shifted forces in molecular dynamics},}\ }\href {\doibase 10.1063/1.3558787}
  {\bibfield  {journal} {\bibinfo  {journal} {The Journal of Chemical Physics}\
  }\textbf {\bibinfo {volume} {134}},\ \bibinfo {pages} {081102} (\bibinfo
  {year} {2011})}\BibitemShut {NoStop}%
\bibitem [{\citenamefont {Widmer-Cooper}, \citenamefont {Harrowell},\ and\
  \citenamefont {Fynewever}(2004)}]{Harrowell2004}%
  \BibitemOpen
  \bibfield  {author} {\bibinfo {author} {\bibfnamefont {A.}~\bibnamefont
  {Widmer-Cooper}}, \bibinfo {author} {\bibfnamefont {P.}~\bibnamefont
  {Harrowell}}, \ and\ \bibinfo {author} {\bibfnamefont {H.}~\bibnamefont
  {Fynewever}},\ }\bibfield  {title} {\enquote {\bibinfo {title} {How
  reproducible are dynamic heterogeneities in a supercooled liquid?}}\ }\href
  {\doibase 10.1103/PhysRevLett.93.135701} {\bibfield  {journal} {\bibinfo
  {journal} {Physical Review Letters}\ }\textbf {\bibinfo {volume} {93}},\
  \bibinfo {pages} {135701} (\bibinfo {year} {2004})}\BibitemShut {NoStop}%
\bibitem [{\citenamefont {Sengupta}\ \emph {et~al.}(2013)\citenamefont
  {Sengupta}, \citenamefont {Karmakar}, \citenamefont {Dasgupta},\ and\
  \citenamefont {Sastry}}]{Sengupta2013}%
  \BibitemOpen
  \bibfield  {author} {\bibinfo {author} {\bibfnamefont {S.}~\bibnamefont
  {Sengupta}}, \bibinfo {author} {\bibfnamefont {S.}~\bibnamefont {Karmakar}},
  \bibinfo {author} {\bibfnamefont {C.}~\bibnamefont {Dasgupta}}, \ and\
  \bibinfo {author} {\bibfnamefont {S.}~\bibnamefont {Sastry}},\ }\bibfield
  {title} {\enquote {\bibinfo {title} {{Breakdown of the Stokes-Einstein
  relation in two, three and four dimensions}},}\ }\href {\doibase
  10.1063/1.4792356} {\bibfield  {journal} {\bibinfo  {journal} {J. Chem.
  Phys.}\ }\textbf {\bibinfo {volume} {138}},\ \bibinfo {pages} {12A548}
  (\bibinfo {year} {2013})}\BibitemShut {NoStop}%
\bibitem [{\citenamefont {Shiba}, \citenamefont {Kawasaki},\ and\ \citenamefont
  {Kim}(2019)}]{Shiba2019}%
  \BibitemOpen
  \bibfield  {author} {\bibinfo {author} {\bibfnamefont {H.}~\bibnamefont
  {Shiba}}, \bibinfo {author} {\bibfnamefont {T.}~\bibnamefont {Kawasaki}}, \
  and\ \bibinfo {author} {\bibfnamefont {K.}~\bibnamefont {Kim}},\ }\bibfield
  {title} {\enquote {\bibinfo {title} {Local density fluctuation governs
  divergence of viscosity underlying elastic and hydrodynamic anomalies in a 2d
  glass-forming liquid},}\ }\href {http://arxiv.org/abs/1905.05458} {\bibfield
  {journal} {\bibinfo  {journal} {Physical Review Letters}\ }\textbf {\bibinfo
  {volume} {123}},\ \bibinfo {pages} {265501} (\bibinfo {year}
  {2019})}\BibitemShut {NoStop}%
\bibitem [{\citenamefont {Mizuno}, \citenamefont {Shiba},\ and\ \citenamefont
  {Ikeda}(2017)}]{Mizuno2017}%
  \BibitemOpen
  \bibfield  {author} {\bibinfo {author} {\bibfnamefont {H.}~\bibnamefont
  {Mizuno}}, \bibinfo {author} {\bibfnamefont {H.}~\bibnamefont {Shiba}}, \
  and\ \bibinfo {author} {\bibfnamefont {A.}~\bibnamefont {Ikeda}},\ }\bibfield
   {title} {\enquote {\bibinfo {title} {Continuum limit of the vibrational
  properties of amorphous solids},}\ }\href {\doibase 10.1073/pnas.1709015114}
  {\bibfield  {journal} {\bibinfo  {journal} {Proceedings of the National
  Academy of Sciences of the United States of America}\ }\textbf {\bibinfo
  {volume} {114}},\ \bibinfo {pages} {E9767--E9774} (\bibinfo {year}
  {2017})}\BibitemShut {NoStop}%
\bibitem [{\citenamefont {Richard}\ \emph {et~al.}(2021)\citenamefont
  {Richard}, \citenamefont {Kapteijns}, \citenamefont {Giannini}, \citenamefont
  {Manning},\ and\ \citenamefont {Lerner}}]{Manning2021}%
  \BibitemOpen
  \bibfield  {author} {\bibinfo {author} {\bibfnamefont {D.}~\bibnamefont
  {Richard}}, \bibinfo {author} {\bibfnamefont {G.}~\bibnamefont {Kapteijns}},
  \bibinfo {author} {\bibfnamefont {J.~A.}\ \bibnamefont {Giannini}}, \bibinfo
  {author} {\bibfnamefont {M.~L.}\ \bibnamefont {Manning}}, \ and\ \bibinfo
  {author} {\bibfnamefont {E.}~\bibnamefont {Lerner}},\ }\bibfield  {title}
  {\enquote {\bibinfo {title} {Simple and broadly applicable definition of
  shear transformation zones},}\ }\href {\doibase
  10.1103/PhysRevLett.126.015501} {\bibfield  {journal} {\bibinfo  {journal}
  {Physical Review Letters}\ }\textbf {\bibinfo {volume} {126}},\ \bibinfo
  {pages} {015501} (\bibinfo {year} {2021})}\BibitemShut {NoStop}%
\bibitem [{\citenamefont {Hu}\ and\ \citenamefont {Tanaka}(2022)}]{Tanaka2022}%
  \BibitemOpen
  \bibfield  {author} {\bibinfo {author} {\bibfnamefont {Y.-C.}\ \bibnamefont
  {Hu}}\ and\ \bibinfo {author} {\bibfnamefont {H.}~\bibnamefont {Tanaka}},\
  }\bibfield  {title} {\enquote {\bibinfo {title} {Origin of the boson peak in
  amorphous solids},}\ }\href {\doibase 10.1038/s41567-022-01628-6} {\bibfield
  {journal} {\bibinfo  {journal} {Nature Physics}\ }\textbf {\bibinfo {volume}
  {18}},\ \bibinfo {pages} {669--677} (\bibinfo {year} {2022})}\BibitemShut
  {NoStop}%
\bibitem [{\citenamefont {Yunker}\ \emph {et~al.}(2009)\citenamefont {Yunker},
  \citenamefont {Zhang}, \citenamefont {Aptowicz},\ and\ \citenamefont
  {Yodh}}]{Yunker2009}%
  \BibitemOpen
  \bibfield  {author} {\bibinfo {author} {\bibfnamefont {P.}~\bibnamefont
  {Yunker}}, \bibinfo {author} {\bibfnamefont {Z.}~\bibnamefont {Zhang}},
  \bibinfo {author} {\bibfnamefont {K.~B.}\ \bibnamefont {Aptowicz}}, \ and\
  \bibinfo {author} {\bibfnamefont {A.~G.}\ \bibnamefont {Yodh}},\ }\bibfield
  {title} {\enquote {\bibinfo {title} {Irreversible rearrangements, correlated
  domains, and local structure in aging glasses},}\ }\href {\doibase
  10.1103/PhysRevLett.103.115701} {\bibfield  {journal} {\bibinfo  {journal}
  {Physical Review Letters}\ }\textbf {\bibinfo {volume} {103}},\ \bibinfo
  {pages} {115701} (\bibinfo {year} {2009})}\BibitemShut {NoStop}%
\bibitem [{\citenamefont {Chikkadi}\ and\ \citenamefont
  {Schall}(2012)}]{Schall2012}%
  \BibitemOpen
  \bibfield  {author} {\bibinfo {author} {\bibfnamefont {V.}~\bibnamefont
  {Chikkadi}}\ and\ \bibinfo {author} {\bibfnamefont {P.}~\bibnamefont
  {Schall}},\ }\bibfield  {title} {\enquote {\bibinfo {title} {Nonaffine
  measures of particle displacements in sheared colloidal glasses},}\ }\href
  {\doibase 10.1103/PhysRevE.85.031402} {\bibfield  {journal} {\bibinfo
  {journal} {Physical Review E}\ }\textbf {\bibinfo {volume} {85}},\ \bibinfo
  {pages} {031402} (\bibinfo {year} {2012})}\BibitemShut {NoStop}%
\bibitem [{\citenamefont {Shiba}, \citenamefont {Kawasaki},\ and\ \citenamefont
  {Onuki}(2012)}]{Shiba2012}%
  \BibitemOpen
  \bibfield  {author} {\bibinfo {author} {\bibfnamefont {H.}~\bibnamefont
  {Shiba}}, \bibinfo {author} {\bibfnamefont {T.}~\bibnamefont {Kawasaki}}, \
  and\ \bibinfo {author} {\bibfnamefont {A.}~\bibnamefont {Onuki}},\ }\bibfield
   {title} {\enquote {\bibinfo {title} {Relationship between bond-breakage
  correlations and four-point correlations in heterogeneous glassy dynamics:
  Configuration changes and vibration modes},}\ }\href@noop {} {\bibfield
  {journal} {\bibinfo  {journal} {Physical Review E}\ }\textbf {\bibinfo
  {volume} {86}},\ \bibinfo {pages} {041504} (\bibinfo {year}
  {2012})}\BibitemShut {NoStop}%
\bibitem [{\citenamefont {Guiselin}, \citenamefont {Scalliet},\ and\
  \citenamefont {Berthier}(2022)}]{Guiselin2022}%
  \BibitemOpen
  \bibfield  {author} {\bibinfo {author} {\bibfnamefont {B.}~\bibnamefont
  {Guiselin}}, \bibinfo {author} {\bibfnamefont {C.}~\bibnamefont {Scalliet}},
  \ and\ \bibinfo {author} {\bibfnamefont {L.}~\bibnamefont {Berthier}},\
  }\bibfield  {title} {\enquote {\bibinfo {title} {Microscopic origin of excess
  wings in relaxation spectra of supercooled liquids},}\ }\href {\doibase
  10.1038/s41567-022-01508-z} {\bibfield  {journal} {\bibinfo  {journal}
  {Nature Physics}\ }\textbf {\bibinfo {volume} {18}},\ \bibinfo {pages}
  {468--472} (\bibinfo {year} {2022})}\BibitemShut {NoStop}%
\bibitem [{\citenamefont {Falk}\ and\ \citenamefont {Langer}(1998)}]{Falk1998}%
  \BibitemOpen
  \bibfield  {author} {\bibinfo {author} {\bibfnamefont {M.~L.}\ \bibnamefont
  {Falk}}\ and\ \bibinfo {author} {\bibfnamefont {J.~S.}\ \bibnamefont
  {Langer}},\ }\bibfield  {title} {\enquote {\bibinfo {title} {Dynamics of
  viscoplastic deformation in amorphous solids},}\ }\href@noop {} {\bibfield
  {journal} {\bibinfo  {journal} {Physical Review E}\ }\textbf {\bibinfo
  {volume} {57}},\ \bibinfo {pages} {7192--7205} (\bibinfo {year}
  {1998})}\BibitemShut {NoStop}%
\bibitem [{\citenamefont {Shimizu}, \citenamefont {Ogata},\ and\ \citenamefont
  {Li}(2007)}]{2007Ogata}%
  \BibitemOpen
  \bibfield  {author} {\bibinfo {author} {\bibfnamefont {F.}~\bibnamefont
  {Shimizu}}, \bibinfo {author} {\bibfnamefont {S.}~\bibnamefont {Ogata}}, \
  and\ \bibinfo {author} {\bibfnamefont {J.}~\bibnamefont {Li}},\ }\bibfield
  {title} {\enquote {\bibinfo {title} {Theory of shear banding in metallic
  glasses and molecular dynamics calculations},}\ }\href {\doibase
  10.2320/matertrans.MJ200769} {\bibfield  {journal} {\bibinfo  {journal}
  {Materials Transactions}\ }\textbf {\bibinfo {volume} {48}},\ \bibinfo
  {pages} {2923--2927} (\bibinfo {year} {2007})}\BibitemShut {NoStop}%
\bibitem [{\citenamefont {Hasyim}\ and\ \citenamefont
  {Mandadapu}(2021)}]{Mandaptu2022}%
  \BibitemOpen
  \bibfield  {author} {\bibinfo {author} {\bibfnamefont {M.~R.}\ \bibnamefont
  {Hasyim}}\ and\ \bibinfo {author} {\bibfnamefont {K.~K.}\ \bibnamefont
  {Mandadapu}},\ }\bibfield  {title} {\enquote {\bibinfo {title} {A theory of
  localized excitations in supercooled liquids},}\ }\href {\doibase
  10.1063/5.0056303} {\bibfield  {journal} {\bibinfo  {journal} {Journal of
  Chemical Physics}\ }\textbf {\bibinfo {volume} {155}},\ \bibinfo {pages}
  {044504} (\bibinfo {year} {2021})}\BibitemShut {NoStop}%
\bibitem [{\citenamefont {Shiba}\ \emph {et~al.}(2016)\citenamefont {Shiba},
  \citenamefont {Yamada}, \citenamefont {Kawasaki},\ and\ \citenamefont
  {Kim}}]{Shiba2016}%
  \BibitemOpen
  \bibfield  {author} {\bibinfo {author} {\bibfnamefont {H.}~\bibnamefont
  {Shiba}}, \bibinfo {author} {\bibfnamefont {Y.}~\bibnamefont {Yamada}},
  \bibinfo {author} {\bibfnamefont {T.}~\bibnamefont {Kawasaki}}, \ and\
  \bibinfo {author} {\bibfnamefont {K.}~\bibnamefont {Kim}},\ }\bibfield
  {title} {\enquote {\bibinfo {title} {Unveiling dimensionality dependence of
  glassy dynamics: 2d infinite fluctuation eclipses inherent structural
  relaxation},}\ }\href {\doibase 10.1103/PhysRevLett.117.245701} {\bibfield
  {journal} {\bibinfo  {journal} {Physical Review Letters}\ }\textbf {\bibinfo
  {volume} {117}},\ \bibinfo {pages} {245701} (\bibinfo {year}
  {2016})}\BibitemShut {NoStop}%
\bibitem [{\citenamefont {Peng}, \citenamefont {Li},\ and\ \citenamefont
  {Wang}(2011)}]{Peng2011}%
  \BibitemOpen
  \bibfield  {author} {\bibinfo {author} {\bibfnamefont {H.~L.}\ \bibnamefont
  {Peng}}, \bibinfo {author} {\bibfnamefont {M.~Z.}\ \bibnamefont {Li}}, \ and\
  \bibinfo {author} {\bibfnamefont {W.~H.}\ \bibnamefont {Wang}},\ }\bibfield
  {title} {\enquote {\bibinfo {title} {Structural signature of plastic
  deformation in metallic glasses},}\ }\href {\doibase
  10.1103/PhysRevLett.106.135503} {\bibfield  {journal} {\bibinfo  {journal}
  {Physical Review Letters}\ }\textbf {\bibinfo {volume} {106}},\ \bibinfo
  {pages} {135503} (\bibinfo {year} {2011})}\BibitemShut {NoStop}%
\bibitem [{\citenamefont {Jung}, \citenamefont {Biroli},\ and\ \citenamefont
  {Berthier}(2022)}]{Jung2022}%
  \BibitemOpen
  \bibfield  {author} {\bibinfo {author} {\bibfnamefont {G.}~\bibnamefont
  {Jung}}, \bibinfo {author} {\bibfnamefont {G.}~\bibnamefont {Biroli}}, \ and\
  \bibinfo {author} {\bibfnamefont {L.}~\bibnamefont {Berthier}},\ }\href
  {http://arxiv.org/abs/2210.16623} {\enquote {\bibinfo {title} {Predicting
  dynamic heterogeneity in glass-forming liquids by physics-informed machine
  learning},}\ } (\bibinfo {year} {2022}),\ \Eprint
  {http://arxiv.org/abs/2210.16623} {arXiv:2210.16623} \BibitemShut {NoStop}%
\bibitem [{\citenamefont {Pezzicoli}, \citenamefont {Charpiat},\ and\
  \citenamefont {Landes}(2022)}]{Pezzicoli2022}%
  \BibitemOpen
  \bibfield  {author} {\bibinfo {author} {\bibfnamefont {F.~S.}\ \bibnamefont
  {Pezzicoli}}, \bibinfo {author} {\bibfnamefont {G.}~\bibnamefont {Charpiat}},
  \ and\ \bibinfo {author} {\bibfnamefont {F.~P.}\ \bibnamefont {Landes}},\
  }\href {http://arxiv.org/abs/2211.03226} {\enquote {\bibinfo {title}
  {Se(3)-equivariant graph neural networks for learning glassy liquids
  representations},}\ } (\bibinfo {year} {2022}),\ \Eprint
  {http://arxiv.org/abs/2211.03226} {arXiv:2211.03226} \BibitemShut {NoStop}%
\bibitem [{\citenamefont {Jiang}, \citenamefont {Tian},\ and\ \citenamefont
  {Li}(2022)}]{Jiang2022}%
  \BibitemOpen
  \bibfield  {author} {\bibinfo {author} {\bibfnamefont {X.}~\bibnamefont
  {Jiang}}, \bibinfo {author} {\bibfnamefont {Z.}~\bibnamefont {Tian}}, \ and\
  \bibinfo {author} {\bibfnamefont {K.}~\bibnamefont {Li}},\ }\href
  {http://arxiv.org/abs/2211.12832} {\enquote {\bibinfo {title}
  {Geometry-enhanced graph neural network for glassy dynamics prediction},}\ }
  (\bibinfo {year} {2022}),\ \Eprint {http://arxiv.org/abs/2211.12832}
  {arXiv:2211.12832} \BibitemShut {NoStop}%
\bibitem [{\citenamefont {Noé}\ \emph {et~al.}(2019)\citenamefont {Noé},
  \citenamefont {Olsson}, \citenamefont {Köhler},\ and\ \citenamefont
  {Wu}}]{Noe2019}%
  \BibitemOpen
  \bibfield  {author} {\bibinfo {author} {\bibfnamefont {F.}~\bibnamefont
  {Noé}}, \bibinfo {author} {\bibfnamefont {S.}~\bibnamefont {Olsson}},
  \bibinfo {author} {\bibfnamefont {J.}~\bibnamefont {Köhler}}, \ and\
  \bibinfo {author} {\bibfnamefont {H.}~\bibnamefont {Wu}},\ }\bibfield
  {title} {\enquote {\bibinfo {title} {Boltzmann generators: Sampling
  equilibrium states of many-body systems with deep learning},}\ }\href
  {\doibase 10.1126/science.aaw1147} {\bibfield  {journal} {\bibinfo  {journal}
  {Science}\ }\textbf {\bibinfo {volume} {365}},\ \bibinfo {pages} {eaaw1147}
  (\bibinfo {year} {2019})}\BibitemShut {NoStop}%
\bibitem [{\citenamefont {Gabrié}, \citenamefont {Rotskoff},\ and\
  \citenamefont {Vanden-Eijnden}(2022)}]{Gabrie2022}%
  \BibitemOpen
  \bibfield  {author} {\bibinfo {author} {\bibfnamefont {M.}~\bibnamefont
  {Gabrié}}, \bibinfo {author} {\bibfnamefont {G.~M.}\ \bibnamefont
  {Rotskoff}}, \ and\ \bibinfo {author} {\bibfnamefont {E.}~\bibnamefont
  {Vanden-Eijnden}},\ }\bibfield  {title} {\enquote {\bibinfo {title} {Adaptive
  monte carlo augmented with normalizing flows},}\ }\href {\doibase
  10.1073/pnas.2109420119/-/DCSupplemental} {\bibfield  {journal} {\bibinfo
  {journal} {Proceedings of the National Academy of Sciences of the United
  States of America}\ }\textbf {\bibinfo {volume} {119}},\ \bibinfo {pages}
  {e2109420119} (\bibinfo {year} {2022})}\BibitemShut {NoStop}%
\bibitem [{\citenamefont {Fey}\ and\ \citenamefont
  {Lenssen}(2019)}]{Fey/Lenssen/2019}%
  \BibitemOpen
  \bibfield  {author} {\bibinfo {author} {\bibfnamefont {M.}~\bibnamefont
  {Fey}}\ and\ \bibinfo {author} {\bibfnamefont {J.~E.}\ \bibnamefont
  {Lenssen}},\ }\bibfield  {title} {\enquote {\bibinfo {title} {Fast graph
  representation learning with {PyTorch Geometric}},}\ }in\ \href@noop {}
  {\emph {\bibinfo {booktitle} {ICLR Workshop on Representation Learning on
  Graphs and Manifolds}}}\ (\bibinfo {year} {2019})\BibitemShut {NoStop}%
\end{thebibliography}%
\end{document}